06/09/2006

# Power laws in microrheology experiments on living cells: comparative analysis and modelling

Martial Balland[1§], Nicolas Desprat[1§], Delphine Icard[1], Sophie Féréol[2], Atef Asnacios[1], Julien Browaeys[1], Sylvie Hénon[1] and François Gallet[1]

[1] Laboratoire Matière et Systèmes Complexes
UMR 7057 associée au CNRS et à l'Université Paris7 - Denis Diderot
Case courrier 7056 – 2, place Jussieu
75251 Paris cedex 05 – France

[2] Physiopathologie et Thérapeutique Respiratoires
INSERM U492 – Faculté de Médecine et Faculté des Sciences et Technologie
8, rue du Général Sarrail
94010 Creteil Cedex - France

[§] These two authors equally contributed to this work.

**Abstract**

We compare and synthesize the results of two microrheological experiments on the cytoskeleton of single cells. In the first one, the creep function $J(t)$ of a cell stretched between two glass plates is measured after applying a constant force step. In the second one, a microbead specifically bound to transmembrane receptors is driven by an oscillating optical trap, and the viscoelastic coefficient $G_e(\omega)$ is retrieved. Both $J(t)$ and $G_e(\omega)$ exhibit power law behaviors: $J(t) = A_0(\frac{t}{t_0})^\alpha$ and $\left|G_e(\omega)\right| = G_0(\frac{\omega}{\omega_0})^\alpha$, with the same exponent $\alpha \approx 0.2$. This power law behavior is very robust ; $\alpha$ is distributed over a narrow range, and shows almost no dependance on the cell type, on the nature of the protein complex which transmits the mechanical stress, nor on the typical length scale of the experiment. On the contrary, the prefactors $A_0$ and $G_0$ appear very sensitive to these parameters. Whereas the exponents $\alpha$ are normally distributed over the cell population, the prefactors $A_0$ and $G_0$ follow a log-normal repartition. These results are compared with other data published in the litterature. We propose a global interpretation, based on a semi-phenomenological model, which involves a broad distribution of relaxation times in the system. The model predicts the power law behavior and the statistical repartition of the mechanical parameters, as experimentally observed for the cells. Moreover, it leads to an estimate of the largest response time in the cytoskeletal network: $\tau_m \sim 1000$ s.

# I - Introduction

To perform their functions, living cells must adapt to external stresses and to varying mechanical properties of their environment. Thus, rheological properties (*i.e.* stress-strain relationships) are key features of living cells. Actually, mechanics play a major role in many biological processes such as cell crawling, wound healing, protein regulation and even apoptosis [1]. Conversely, several pathologies, like metastasis, asthma or sickle cell anemia, involve alteration of the mechanical properties of a given cell type. All these processes are mainly controlled by the structure and mechanical properties of the cytoskeletal network. This network is a dynamical assembly of macromolecules, principally made of actin filaments, intermediate filaments and microtubules, and interacting with a variety of associated proteins, crosslinkers, and molecular motors. The mechanical properties of the cytoskeletal network are therefore the subject of many experimental studies.

These studies have been made possible by the development of numerous quantitative micromanipulation techniques, such as micropipettes [2], cell poking [3], shear flow cytometry [4, 5], atomic force microscopy (AFM) [6-8], microplates [9-11], optical tweezers [12-14], optical stretchers [15, 16], magnetic tweezers [17, 18], magnetic twisters [12, 19], particle tracking [20-22]. These techniques are complementary, in the sense that they probe the behavior of the intracellular medium at different length scales and time scales, and that they implement stresses and strains in different geometries and with different orders of magnitude.

Most recent results in microrheology of the intracellular medium have demonstrated that it is a complex viscoelastic medium which cannot be simply modelized by associating a finite, small number of elastic and viscous elements. Indeed, the viscoelastic complex modulus of the cell medium exhibits a power law behavior on a wide frequency range [13, 23]. Similarly the creep function behaves as a power law of elapsed time [11, 24]. This clearly indicates that there is a broad and dense distribution of dissipation times in the cell, and that the mechanisms responsible for the storage of elastic energy and its dissipation are strongly correlated. Such behavior, characteristic of structural damping, is found in other complex viscoelastic systems, like colloids, gels, pastes, and more generally the class of so-called "soft glassy materials". However, a detailed interpretation of the origin of structural damping behavior in the cytoskeletal network has still to be built.

One of the objectives of this paper is to propose a global description of the mechanics of the cytoskeletal network, consistent with the whole set of data gathered from numerous experiments performed up to now. It aims at contributing to answer several questions, such as: how might one compare the results of experiments performed with different microrheological techniques and/or in various experimental conditions ? Is there a unified behavior at the nano- and micro-scale level and at the scale of the whole cell ? How does this behavior depend on the cell type, and on the nature of the receptor through which the stress is applied ?  Is it possible to build a phenomenological model which predicts the power law response functions and also the peculiar dispersion of results obtained from a set of cells of the same type ?

The paper is organized as follows: section II presents the experimental protocols and the two techniques used here to measure either the local dynamical response of the actin network to an oscillatory force (optical tweezers) or the creep function at the scale of the whole cell (uniaxial stretching). The results of the two experiments are reported in section III. The power laws measured for the viscoelastic modulus *vs* frequency and for the creep function *vs* time are fully consistent. The prefactor $G_0$ of the power law, *i.e.* the elastic modulus measured for the same cell type at the reference frequency $f$ = 1Hz, follows a log-normal distribution. A comparison between different cell types and receptors shows that the value of the exponent $\alpha$ in the power law is remarkably homogeneous: $\alpha \approx 0.2$. Section IV presents a review of other data collected in the litterature. The common features emerging from the different cell types, probed by different techniques and in various experimental conditions are underlined. The possible origins for some discrepancies, mainly in the prefactor of the power laws, are discussed. An original, semi-phenomenological model for the cell mechanical behavior is set out in section V. We assume that the cell is built of a random infinite series of elementary mechanical units, following on average a scaling law repartition. It leads to correct predictions for the response functions, including power law behaviors, normal and log-normal distributions respectively for the exponent $\alpha$ and the prefactor $G_0$. Section VI details the quantitative comparison with experimental data, leading to an estimate for the largest relaxation time in the cell $\tau_m$.

## II – Experimental setups and protocols

*- Cell culture*

We studied the microrheology of two primary cell cultures, mouse ear fibroblasts and rat alveolar macrophages, and four different cell lines: C2 and C2-7 myogenic cells, A549 human alveolar epithelial cells, Madin–Darby Canine Kidney (MDCK) epithelial cells, and mouse fibroblasts L929.

The C2 myogenic cell line is derived from the skeletal muscle of adult CH3 mice. The C2-7 cell line is a subclone of the C2 cell line. They were kindly provided by M. Lambert and R.M. Mège (INSERM U440, Institut du Fer à Moulin, Paris) and by D. Paulin (Biologie Moléculaire de la Différentiation EA 300, Université Paris VII, Paris). A549 are human lung carcinoma cells (American Type Culture Collection, Rockville, MD). MDCK cells were kindly provided by Ph. Chavrier (Membrane and Cytoskeleton Dynamics, CNRS, Institut Curie, Paris). Alveolar macrophages were isolated from Sprague-Dawley rats by broncho-alveolar lavages and re-suspended in RPMI medium supplemented with 0.1% BSA [25]. Other cells were grown at 37°C in a humidified 5% $CO_2$ - 95% air atmosphere, in Dulbecco's Modified Essential Medium supplemented with 10% fetal calf serum, 2 mM glutamine, 100 units/ml penicillin and 50 mg/ml streptomycin.

For creep experiments, the cells were detached from culture flasks (1% trypsin and 1 mM EDTA), centrifugated at 900 rpm for 3 min, diluted in DMEM supplemented with 15mM HEPES, and maintained under smooth agitation for 2 h at 37°C.

For optical tweezers experiments, the cells were detached from culture flasks 24 hours before experiments and plated at a density of about 300 cells/mm$^2$, in complete culture medium with serum, on glass coverslips (22 mm × 22 mm) coated with fibronectin (5 µg/mL for 3h at room temperature). For optical tweezers experiments on A549 *via* ICAM-1, 10 ng/ml of recombinant human Tumor Necrosis Factor-α was added to this medium to induce the expression of ICAM-1.

*- Microplates preparation*

Glass microplates were cleaned for 10 min in a Piranha mixture - 2/3 pure sulfuric acid, 1/3 hydrogen peroxide 30% and rinsed in water.

For non-specific binding, microplates were dipped in a bath of 90% ethanol, 8% water, 2% 3-aminopropyltriethoxysilane for 2 h, rinsed in ethanol, and finally incubated in 98% water, 2% glutaraldehyde 1 h before the experiment.

For specific binding to integrins, microplates were dipped in a bath of 5*mL* DMEM at *1mg/mL* fibronectine F-1141 (Sigma), 2 h before the experiment.

For specific binding to cadherins, microplates were directly coated in the experimental chamber to avoid drying of the proteins. The microplates tips, previously coated with organopolysiloxane (Sigmacote, Sigma) [26], were first dipped, for 2 h, in a 150 *µL* drop of borate buffer containing 2 *µL* of a 2 *µg/µL* Fcγ antibody solution (Jackson ImmunoResearch, West Grove, PA, USA) [27]. After rinsing with borate buffer, microplates were dipped for 2 h in a 200 $\mu L$ borate buffer containing 25 $\mu g$ of NCad-Fc "chimeric" protein. Then the plates tips were rinsed and finally dipped in a 2 *mL* borate buffer containing 10 *mg/mL* BSA (Sigma-Aldrich, France), to ensure saturation of residual non-specific adhesion sites.

*- Bead coatings and specific attachment to the cells*

For a specific binding to integrins, carboxylated silica microbeads (3.47 µm diameter, Bangs Laboratories Inc., IN, USA) were coated with a polypeptide containing the arginine–glycine–aspartic (RGD) sequence (PepTite-2000, Telios pharmaceuticals, CA, USA), according to the manufacturer's procedure.

For specific binding to ICAM-1 (CD54), the same silica beads were first coated with goat anti-mouse IgG (BD Biosciences, 5µg of monoclonal antibody for 100µg of beads, in PBS for 30 min at 4°C under gentle agitation), and secondly with mouse anti-human CD54 (BD Biosciences, same protocol).

Before use, coated beads were incubated in DMEM supplemented with 1% BSA for 30 minutes at 37°C, to block non-specific binding. Beads were then added to the cells (~ 5µg of beads per coverslip) and further incubated for 30 min at 37°C. Unbound beads were washed away with medium.

*- Measurements of the creep function J(t)*

The creep function of a single living cell is measured by means of a home-made stretching rheometer, which has been described in detail elsewhere [11, 28]. This setup takes advantage of a simple uniaxial geometry, since the cell is stretched between two glass microplates, a rigid one and a flexible one [9]. The two arms bearing the microplates are set up on each side of the inverted microscope, perpendicularly to the optical axis. A XYZ piezoelectric stage, interfaced with a computer, allows to accurately control the gap between the plates. The stiffness of the flexible plate is independantly calibrated, therefore it is possible to simultaneously measure the force applied to the cell, and its deformation. A real time detection and an efficient feedback loop were implemented, in order to monitor and control the flexible plate deflection. Thus the setup can be used as a constant stress micro-rheometer. The whole setup is isolated from vibrations and is maintained at 37 ± 0.2 °C.

Each cell stretched in the gap between the microplates is visualized under bright light illumination and its image is recorded with a digital CCD camera, as shown for instance on Figure 1. The position of the tip of the thin flexible plate is held constant to ensure that the plate deflection and therefore the applied force remain constant. The strain $\varepsilon(t)$ is defined as $(L(t)-L_0)/L_0$, where $L(t)$ represents the cell length at time $t$ and $L_0 = L(0)$ its initial length when the force step is applied. The applied stress $\sigma_0$ is taken as the ratio between the constant applied force $F_0$ and an effective area $A$ of the cell section. An estimate of this effective area is given by the mean value of the measured contact areas of the cell at each microplate, assumed to be circular. For each cell, the creep function $J(t)$ is derived from:

$$J(t) = \varepsilon(t)/\sigma_0 \quad (1).$$

*- Measurement of the complex viscoelastic modulus $G_e(\omega)$.*

An optical tweezers setup is used to measure the viscoelastic modulus of the cytoskeletal network, by applying an oscillating force to a microbead specifically bound to transmembrane receptors (see cell culture and coating). The setup and force calibration have been previously described [13, 29]. The experimental chamber is mounted on a piezoelectric stage and is submitted to a sinusoïdal displacement at frequency $f=\omega/2\pi$, while the trap is kept at a fixed position (Figure 2). The displacements of the chamber $x_c(t)$ and of the bead $x_b(t)$ are recorded with a fast (500 Hz) CCD camera. At any time, the force $F(t) = \hat{F}(\omega)\exp(j\omega t)$ exerted on the bead is calculated from the bead-trap distance $x_b$ according to the calibration curves, while the cellular deformation is related to the relative displacement $x(t) = x_c(t) - x_b(t) = \hat{x}(\omega)\exp(j\omega t)$ between the chamber and the bead. Here $\hat{F}(\omega)$ and $\hat{x}(\omega)$ are complex numbers, with a relative non-zero phase. Using Labview®, we generate a sequence of successive sinusoidal signals at given frequencies (from 0.05 to 50 Hz), which control the piezoelectric ceramic motion $x_c(t)$, and a synchronous sequence of pulses to trigger the image acquisition. In order to minimize a possible actin remodelling at the bead periphery, or the effect of a slow drift of the bead on the cell surface, the total duration of a measurement on the same single cell never exceeds 2.5 min, a time short as compared to the 30 min incubation time for the adherence of the bead to the cell.

The derivation of the complex modulus $G_e(\omega)$ from $\hat{F}(\omega)$ and $\hat{x}(\omega)$ is described in appendix B, and requires formula (B5):

$$\hat{F}(\omega) = 2\pi R G_e(\omega) f(\Theta) \hat{x}(\omega) \quad (2)$$

Here the cell is considered as an elastic homogeneous medium. The dimensionless factor $f(\Theta)$, for which we calculated an analytic expression (formula B3) [12], accounts for the actual immersion of the bead into the cell. The parameter $\Theta$ is defined as half the angle of the immersion cone of the bead or radius R into the medium. Since we restrict our measurements to the beads bound to the side of the cells, $\Theta$ is estimated from axial image recordings in the vertical direction (Figure 3).

## III - Measurements of the viscoelastic parameters for different cell types.

*- Creep function and viscoelastic modulus of premuscular C2 cells*

A typical recording of the creep function $J(t)$, obtained for a single C2 cell with the stretching rheometer, is shown in figure 4. Over three orders of magnitude in time (0.1 < t <100 s), $J(t)$ is remarkably well fitted by a power law $J(t) = A_0(\frac{t}{t_0})^{\alpha}$ , with $\alpha$ = 0.25 and $A_0$ = 0.017 Pa$^{-1}$. We stress on the fact that $t_0$ is not a parameter of the fit, but an arbitrary reference time, for which $J(t)$ takes the value $A_0$. The dimension of $A_0$ is therefore the same as $J$. For all the measurements, we chose for convenience $t_0$ = 1s. As already shown in [11], any attempt to adjust $J(t)$ by a sum of a few exponential functions in the full time range leads to a much poorer agreement. There is an exact equivalence between a power law behavior of $J(t)$ as a function of time $t$, and a power law behavior of the complex viscoelastic modulus $G_e(\omega)$ as a function of the frequency $\omega$ (equations (A5-A7) in appendix A). Consequently, the value of $|G_e(\omega)|$ derived from this experiment may be expressed as:

$$\left|G_e(\omega)\right| = \frac{\omega^{\alpha} t_0^{\alpha}}{A_0 \Gamma(1+\alpha)} = G_0(\frac{\omega}{\omega_0})^{\alpha} \qquad (3)$$

where $\Gamma$ represents the Euler function (defined in the appendix). From figure 4, one can calculate a typical value $G_0$ = $|G_e(\omega_0)|$ = 103 Pa for this particular curve at the chosen reference frequency $\omega_0/2\pi = 1/t_0$ = 1Hz.

A remarkable feature is that all the tested cells belonging to the same C2 population exhibit a similar power law behavior. Figures 5a and 5b show respectively the measured distributions of the exponent $\alpha$ and of the logarithm $ln(G_0)$ from a set of 43 different C2 cells, glued to the glass plate through a glutaraldehyde coating. On the same figures 5a and 5b are also plotted the cumulative distributions of $\alpha$ and $ln(G_0)$. The cumulative distributions are well adjusted by an error function:

$$E(x) = \frac{1}{2} + \frac{1}{2} erf\left(\frac{x - x_0}{\sqrt{2}\sigma}\right) = \frac{1}{2} + \frac{1}{\sigma\sqrt{2\pi}} \int_0^x exp\left(-\frac{(x'-x_0)^2}{2\sigma^2}\right) dx' \qquad (4)$$

One notices that the distribution of exponents $\alpha$ closely follows a normal law, while the distribution of prefactors $G_0$ appears log-normal. From the erf fits one retrieves the best estimates of the average exponent of the power law $<\alpha>$ = 0.242, of the standard deviation $\sigma_\alpha$ = 0.082, and of the standard error $\Delta_\alpha$ = 0.013 (see table 1). Similarly, for $ln(G_0)$, the estimated averaged value is $<ln(G_0)>$ = 6.46, the standard deviation is $\sigma_G$ = 0.82, and the standard error is $\Delta_G$ = 0.12. Accordingly, for $|G_e(\omega)|$ at 1Hz, one infers the median value $G_{0M}$ = $exp(<ln(G_0)>)$ = 640 (+80/-70) Pa. Notice that the mean value $<G_0>$ = 1060 ± 190 Pa is higher than $G_{0M}$, which is consistent with a log-normal, non symmetric distribution of $G_0$.

A parallel analysis was performed on the same C2 cell type studied with the optical tweezers setup. In this case, by applying a sequence of oscillating stresses to a single cell at different frequencies, one directly measures the complex viscoelastic coefficient $G_e(\omega)$ as a function of $\omega$. In the experiments described below, all the beads are RGD coated and bind to the cells through integrins. In figure 6 the measured values of the modulus $|G_e|$ and of the phase $\delta$ of $G_e(\omega)$ are shown for a single C2 cell, as a function of the frequency $f=\omega/2\pi$, in the range 0.05-50 Hz. As in the cell stretching experiment, $|G_e|$ behaves as a power law of $f$ over three frequency decades. For this particular cell, the exponent $\alpha$ is found equal to 0.30, and the value $G_0$ of $|G_e|$ at 1Hz is found equal to 155 Pa. Moreover, the measured phase shift $\delta$ remains constant, within a good approximation, in the studied frequency range. Its average value is equal to 0.45, very close to the theoretically expected value $\alpha\pi/2$ = 0.47 (see formula A7 in appendix A).

For a whole set of C2 cells tested with optical tweezers, $G_e(\omega)$ present similar power law behaviors. Figure 7a and 7b show the distributions of the exponent $\alpha$ and of the logarithm $ln(G_0)$ of $|G_e|$ at 1Hz, measured for 22 different C2 cells, together with the cumulative distributions of $\alpha$ and $ln(G_0)$. As in figures 5a and 5b, the $\alpha$ distribution follows a normal law, while the distribution of $G_0$ appears log-normal. We have obtained the best estimates: $<\alpha>$ = 0.208 ± 0.021, $G_{0M}$ = 310 (+130/-100) Pa (median value) and $<G_0>$ = 570 ± 150 Pa (mean value).

It is remarkable that the two experiments (uniform stretching and local oscillating force), performed in quite different conditions, lead to the same power law behavior, with approximately the same exponent $\alpha \approx 0.22$ (within experimental error). However, one notices that the prefactor $G_0$ differs in the two cases. Several possible origins of this difference will be discussed in section IV.

*- Viscoelastic modulus of cells excited through different receptors.*

With the optical tweezers setup, we have also performed oscillating force experiments with alveolar epithelial cells (A549). Here we compared the results obtained by using two different mechanical receptors on which the stress is applied: integrins and ICAM-1. ICAM (InterCellular Adhesion Molecules) are transmembrane proteins which allow for instance the macrophages to adhere and to migrate over the pulmonary epithelium [30]. The beads are either coated with RGD peptide or anti-ICAM-1 ligands. Except for the nature of the receptor, the experimental protocols are identical in the two cases. For each single cell, the observed behavior of the elastic modulus $G_e(\omega)$ is very similar in both series of experiments. It is also similar to the one described above for C2 cells: $|G_e(\omega)|$ behaves as a power law of the excitation frequency, while the phase $\delta$ remains approximately constant. The exponent $\alpha$ of the power law is close to 0.2 in both cases. The results are summarized in table 1 and in figures 8 and 9. Over two sets of cells (N=23 and 19 respectively), one observes normal distributions for the exponent $\alpha$ : $<\alpha> = 0.219 \pm 0.014$ for RGD coating, and $<\alpha> = 0.181 \pm 0.014$ for anti-ICAM-1 coating. The prefactor $G_0$ at $f$=1Hz follows a log-normal distribution, with median values $G_{0M}$ = 420 (+80/-70) Pa (RGD coating) and $G_{0M}$ = 80 (+25/-20) Pa (anti-ICAM-1 coating). While the mean values $<\alpha>$ of the exponent appear very close to each other for both coatings, the median value $G_{0M}$ is found appreciably lower when the stress is applied through ICAM-1 than when it is applied through integrins.

Using the uniaxial rheometer we also compared the creep functions of C2 myoblasts stretched: i) through non-specific mechanical receptors (plates coated with glutaraldehyde that binds any protein of the cell surface - data already detailed in the previous section), ii) through cadherins, the specific proteins of cell-cell adhesion. In both cases, the creep function is a week power law of the time. The two sets of cells ($N_{glu}$=43 and $N_{cad}$=13) shows normal distributions for the exponent $\alpha$ , with $<\alpha_{glu}> = 0.242 \pm 0.013$ and $<\alpha_{cad}> = 0.29 \pm 0.02$. The prefactor $G_0$ at $f$=1Hz follows a log-normal distribution, with median values $G_{0M}$ = 640 (+80/-70) Pa (glutaraldehyde) and $G_{0M}$ = 850 (+75/-65) Pa (cadherin). Unlike the Integrin/ICAM-1 comparison in the previous paragraph, here the median values $G_{0M}$ are very close to each other for both coatings, while the mean value $<\alpha>$ of the exponent is found appreciably higher when the stress is applied through cadherins than through glutaraldehyde. However, in both Integrin/ICAM-1 and Cadherin/Glutaraldehyde comparisons, an increase of $<\alpha>$ is related to an increase of $G_{0M}$. Such a correlation between $<\alpha>$ and $G_{0M}$ values can be understood by considering molecular motor activity for instance [13] and is correctly taken into account by the mechanical model presented in section V.

*- Other cell types.*

As reported in table 1, we have performed microrheological experiments on several other cell types, using either uniaxial stretching or optical tweezers, and in various coating conditions. The individual cell behavior appears strikingly independant of the cell type and of the experimental conditions. When stretching the whole cell, the creep function $J(t)$ is accurately adjusted by a power law function of time $t$. Similarly, in oscillating force experiments, the viscoelastic modulus $G_e(\omega)$ behaves as a power law of the excitation frequency. As seen in table 1, the average exponent $<\alpha>$ of the power law always remains in the range 0.15-0.25, whatever the cell type and function: this holds for premuscular cells (C2 myoblasts), epithelial cells (alveolar A549 and MDCK) , fibroblasts (primary and L929), and macrophages (primary). Although the number of cells tested may not always be high enough to yield an accurate statistic, the prefactor $G_0$ of the complex modulus at 1Hz seems to follow a log-normal distribution. Contrary to what is observed for the exponent $\alpha$, the median value $G_{0M}$ of $G_0$ appears to depend on the cell type and on the experimental conditions (see discussion in following section).

## IV - Comparison with other experimental results

The most prominent feature emerging from the experiments is the robustness of the power law behavior, independent of cells types and experimental conditions, as summarized in Table 1. The average values $<\alpha>$ of the exponent remain very close to 0.2. This result is independant of the biological cell function, of the length scale of the experiment, of the nature of the complex transmitting the stress. Other experiments performed on other cell types, and with different techniques, confirm such universal features of the cell mechanical behavior.

*- Optical Magnetic Twisting Cytometry experiments*

First, Fabry and coworkers [19] have performed optical magnetic twisting cytometry (OMTC) on human airway smooth muscle cells (HASM). The experiment consists in applying an oscillatory torque to a magnetic bead bound to the cell membrane, and optically tracking its displacement. They have shown that the viscoelastic modulus $G_s$ follows a power law of the driving frequency, in the range $10^{-2}$ – 100 Hz. They measured an exponent $<\alpha>$ = 0.204 ± 0.002 in control conditions, at a temperature T = 37°C. This exponent only varies when the cells are treated with stiffening or depolymerizing agents of the cytoskeleton. Still in control conditions, and assuming that the magnetic beads are, on average, embedded by 10% into the cell ($<\Theta> \approx 37°$), they compute a median value of $G_s$ at 1Hz approximately equal to 2000 Pa [31]. However their definition of the viscoelastic coefficient $G_s$, which reduces to a static shear modulus $\mu$ at low frequency, is different from ours, $G_e$, which is defined from the Young modulus $E$. Assuming that the cell medium is incompressible ($E/\mu = 3$), this leads to $G_{0M} \approx 6000$ Pa for HASM cells. Lenormand *et al.* [24] have adapted the OMTC technique, and measured the creep function $J_c(t)$ on the same HASM cell type as Fabry and coworkers [19], by applying a constant step torque on a magnetic bead bound to the cell. As expected, they found that $J_c(t)$ is proportional to $t^{\alpha}$, with $<\alpha> = 0.209 \pm 0.003$. Converted into elastic coefficient, the prefactor has the same order of magnitude as in the oscillatory experiment.

For other cell types studied by OMTC [23], power laws were again observed, with quite identical exponents: $<\alpha> = 0.195 \pm 0.005$ for mouse embryonic carcinoma cells (F9), $<\alpha> = 0.173 \pm 0.005$ for human bronchial epithelial cells λ (HBE), $<\alpha> = 0.200 \pm 0.009$ for mouse macrophages (J744.A) and $<\alpha> = 0.186 \pm 0.008$ for human neutrophils. The associated prefactors $G_{0M}$ at 1Hz could not be easily retrieved, since they depend on the exact angle of immersion $\Theta$ of the bead into the cells, which is not reported. Assuming that the height of immersion is comprised between 10% and 30% of the bead diameter ($37° < \Theta < 66°$), and following the simulations of Mijailovitch [31], $G_{0M}$ falls in the range 130-900 Pa (F9), 1000-8000 Pa (HBE), 1500-10000 Pa (J744.A), and 500-3600 Pa (neutrophils).

*- Atomic Force Microscopy experiments*

By probing cells with the oscillating tip of an Atomic Force Microscope (AFM), Alcaraz *et al.* [8] were also able to measure the frequency dependant viscoelastic modulus $G_s(\omega)=G'(\omega)+G''(\omega)$ for epithelial alveolar cells (A549), in the range 0.1 – 100 Hz, at room temperature (T ≈ 20°C). Consistently with the present work and Fabry's work, they observe that both $G'$ and $G''$ are proportional to $\omega^{\alpha}$, with $<\alpha> = 0.22$. The corresponding mean value of $G_0 = G_e(\omega)=3G_s(\omega)$ at 1Hz is $<G_0> \approx 2200$ Pa. Probing the same cell type (A549) by MTC (Magnetic Twisting Cytometry), in the same frequency range, Trepat *et al.* [32] found power laws with an exponent $<\alpha> = 0.214$ and a median value of the prefactor $G_0$: $G_{0M} \approx 2000$ Pa (this value is highly dependant on the bead immersion, here taken equal to 30% on average). Notice that we have measured the same exponent $<\alpha>$ in our experiments on A549 cells (table 1). However the values of $G_{0M}$ (or $<G_0>$) are about 3-5 folds higher in AFM and MTC than the values measured by optical tweezers and stretching rheometer (see table 1). Possible origins of these disagreements are discussed below.

The same AFM technique was applied to human bronchial epithelial cells (BEAS-2B) [8], which should not noticeably differ from the HBE cells probed by Fabry *et al.* Their results are very similar, leading to $<\alpha>$ = 0.20 and $<G_0>$ = 2400 Pa. Independantly, Puig-de-Morales *et al.* probed the same BEAS-2B cell with MTC, at T = 37°C [33]. They retrieved values of $<\alpha>$ = 0.27 and of $<G_0>$ about one order of magnitude lower than in [32].

*- Active and passive microrheology on embedded probes*

It is of great interest to mention the results of some passive microrheology experiments, consisting in measuring the viscoelastic coefficients of the intracellular medium by following the motion of submicron particles embedded in the cytoplasm. Yamada and coworkers [20] measured the mean square displacement $<\Delta r^2(t)>$, as a function of time, of spherical endogenous granules present in kidney epithelial cells COS7, from which they infered the frequency dependant viscoelastic modulus $G_s(\omega)$. The motion of the granules is clearly subdiffusive, and, although the authors do not explicitly mention this interpretation, the variations of $G_s(\omega)$ are consistent with a power law $|G_s(\omega)| = G_{0s}\omega^{\alpha}$ in the range $1 < \omega < 10^4$ rad/s. The exponent $\alpha$ is of the order of 0.5 in the perinuclear region, and 0.33 in the lamellar region, where the density of actin is higher. The average value of $|G_e(\omega)| = 3|G_s(\omega)|$ at 1 Hz ($\omega = 2\pi$ rad/s) is $<G_0> \approx 70$ Pa in the perinuclear region and $<G_0> \approx 210$ Pa in the lamellar region. In a similar way, Tseng *et al.* [21] recorded the motion of carboxylated microspheres microinjected in 3T3 fibroblasts. It is harder in this case to characterize their subdiffusive motion by a single exponent. However, the measured average compliance of the cytoplasm at a time scale $\tau$ = 0.1s is in the range 0.01-0.03 $Pa^{-1}$, which leads to $<G_0>$ around 100-300 Pa at 1Hz. More recently, Yanai *et al.* [14] used optical tweezers to apply step forces on endogenous granules embedded in human neutrophils, and showed that the response is well represented by a power law of exponent about 0.5. Besides, Lau *et al.* [22] measured the two-point correlation function of the mean-square displacement of embedded particles. They found a power law, consistent with an exponent $\alpha$ close to 0.25 for the complex shear modulus. Actually, a two-point microrheology experiment is expected to be more sensitive to the cytoskeletal deformations than one-point microrheology. This indicates at least that one must be cautious when comparing the results of experiments performed on embedded particles and on probes specifically bound to transmembrane receptors.

In an attempt to synthesize this large amount of experimental work, one observes enough common properties to bring out some general features in the cell mechanical behavior:

- the microrheological behavior of the cell is quite accurately described by power laws. This holds either for active microrheology (viscoelastic modulus, creep function) or for passive microrheology (diffusion of particles). This clearly states that the cell mechanics involves a broad distribution of response times.

- except in particle tracking microrheology experiments, in which the probes are not directly attached to the cytoskeleton, the average exponent of the power law lies in a narrow range: $0.15 < \alpha < 0.25$. This result is independant of the experimental technique, of the probe lengthscale, of the cell type and of the nature of the complex transmitting the stress. Adding drugs which induce the stiffening (contractile agents like thrombin, histamin), or the softening (disrupting agents like cytochalasin D or latrunculin) of the cytoskeleton [23, 32, 34], does not induce a dramatic change of this exponent. Only the inhibition of the acto-myosin activity by blebbistatin may cause $\alpha$ to lower down to about 0.10 [13]. The very robust feature of the cell response can only be explained by some common structural organization of the cytoskeletal network, independant of the length scale and of the biological cell function. In section V we propose to model this by a self similar assembly of elementary mechanical units.

- within a set of cells of the same type, probed by the same technique, we observe that the repartition of the viscoelastic modulus $G_0 = |G_e(\omega)|$ at 1 Hz is asymmetric and clearly follows a log-normal distribution. This distribution is characterized by its median value $G_{0M} = exp(<ln\ (G_0)>)$. Other authors also reported such log-normal repartitions [23, 32, 34]. The model developped in section V provides an interpretation of this log-normal distribution.

- from one cell type to another, but also within a same cell type from one experimental technique to another, measurements of $G_{0M}$ - or of $<G_0>$ for the authors who do not make the distinction between $G_{0M}$ and $<G_0>$ - present a wide dispersity (from about $10^2$ to about $10^4$ Pa). A global and consistent interpretation of the origin of this dispersion appears much more difficult to elaborate. Indeed, several competing factors can be invoked to explain the observed differences, some of them being related to intrinsic biological mechanisms, others to experimental conditions or possible artifacts. We enumerate below some of them:

i) The cell type, associated to a given function, is expected to have a relevant influence on the cell average stiffness. The density of actin in the cell, but also the structure of the actin network should indeed largely determine the cell rigidity. Moreover, the activity of molecular motors and their spatial distribution are known to contribute to the cytoskeleton dynamics, and thus to its mechanical behavior. Actually, it appears difficult to bring out from existing data some mechanical characteristics associated to a given cell type. All the more as other factors do also influence the cell stiffness.

ii) In both OMTC and Optical Tweezers experiments, the determination of $G_0$ is very sensitive to the precise knowledge of the bead immersion angle $\Theta$. This parameter is difficult to measure with a good accuracy from the images of single cells. Besides, taking an average value for $<\Theta>$ may lead to overestimate the contribution of weakly bound beads, and thus to underestimate $<G_0>$ [12, 35]. Moreover, $G_0$ is not calculated in the same manner according to different authors. In this work, $G_e(\omega)$ is derived from formula (B5), where the analytical function $f(\Theta)$ accounts for the $\Theta$ dependance (formula B3). In OMTC experiments [19, 23, 24, 32, 34], the $\Theta$ dependance is included in the factor $\beta$ (formula B6), numerically computed in a finite element model [31]. A comparison limited to the bead rotation contributions in both cases shows that the numerical prefactor in (B5) is about twice the one in (B7). This may partly explain why, for the same cell type, the reported values of $G_{0M}$ are smaller in OT experiments than in OMTC experiments.

iii) It is known that the mechanical properties are not homogeneous inside a same cell, and also depend on the cell activity. For fibroblasts, Kole *et al.* [36] have measured that the stiffness in the lamellipodium, where the actin network is denser, is about four times higher than in the perinuclear region. This difference almost vanishes for quiescent fibroblasts. Besides, for migrating neutrophils, Yanai *et al.* showed that $<G_0>$ is almost 10 times smaller in the leading edge than in the cell body or tail [14]. In this case the relative fluidity of the pseudopod may be related to a cytoplasmic flow, driven by a pressure gradient. For spread cells like HeLa, the viscosity experienced by magnetic endosomes embedded in the cytoplasm is found on average eight times higher in the perinuclear region than further away from the cell center [18]. In this last case the correlation to the cell rigidity is not discussed. Other authors [37] have reported that the average cellular medium rigidity decreases by several orders of magnitude when the size of the mechanical probe increases from a few tens nanometers (AFM tip) to a few microns (micropipettes). Although such observations may depend on the cellular type, and consequently general trends are not easy to derive, these different factors may contribute to the dispersion reported above.

iv) Cell viscoelastic properties are also sensitive to the temperature. Lo and Ferrier [38] performed mechanical tests on osteosarcoma cells, and observed that the average stiffness, estimated from a Kelvin-Voigt model, decreases by a factor 1.5 when the temperature increases from 22°C to 37 °C. This effect probably contributes to explain the difference between the high $G_{0M}$ (or $<G_0>$) values measured on A549 and BEAS-2B cells with AFM at 20°C [8, 32] and the lower ones measured by OT or MTC techniques at 37 °C [this work, 23, 33].

v) Comparing the transmission of the mechanical signals through different kinds of receptors remains largely an open question. On HASM cells probed by MTC, it has been shown that the power law exponent does not significantly vary when the bead is covered by RGD (binding to β1 integrin), vitronectin VN (preferentially binding to β3 integrin), uPA urokinase (indirectly binding to the cytoskeleton via caveolin), or AcLDL (which binds to a non-adhesive site, not linked to the cytoskeleton) [34]. However, in the same conditions, the relative value of $G_{0M}$ decreases by a factor of 4 from RGD to VN, and a factor of 10 from RGD to uPA or AcLDL. In this work, we report a difference in the viscoelastic coefficients measured on the same A549 cells through integrin and ICAM-1 receptors on the one hand, on the same C2 cells through glutaraldehyde and cadherin on the other hand (table 1). However, in all these studies, the amount of coating was never precisely controlled, which may also bias the results (see below point vi). Further works are necessary before drawing conclusions about the influence of the receptor/ligand couple. .

vi) Finally, this analysis must include questionning about the structure of the contact itself, which depends on the amount of molecular bonds, and on the dynamical evolution due to cytoskeletal remodeling. Concerning this last point, it is known that applying a constant stress at a focal contact induces within a few minutes local recruitment of integrin, actin and other proteins involved in the contact assembly, which reinforces the contact [39-41]. A similar effect was observed on adherent junctions [42]. Bursac *et al.* [43] have observed that applying a high amplitude oscillating stress to a magnetic bead bound to the cell makes immediately the contact more compliant, and that the contact strengthens again within a few minutes after releasing the stress. But to our knowledge, there is no report of the evolution of $G_0$ during the spontaneous formation of a bead-cell contact, nor during its reinforcement under a continuous stress. A related concern is to determine how the initial number of molecular bonds at the bead-cell contact affects the mechanical response. The coating protocol of the bead by the ligand is obviously a key step, but no method has been yet reported to quantify the amount of ligand in a reliable way.

The above discussion points out the existence of numerous physical and biological parameters, which may interfere in opposite ways, and probably induce the large dispersion of the measured values of the parameter $G_0$. In the absence of more selective experiments focusing on the influence of each parameter, it is yet not possible to go further into a detailed interpretation of the observed differences.

## V - A model for the rheological behavior of a single cell

There are numerous examples in the literature of complex viscoelastic systems showing power law rheological behaviors. This is for instance the case of colloidal systems close to the sol/gel transition [44], or of "soft glassy materials", which includes foams, pastes, emulsions and slurries [45, 46]. A common feature of all these systems is that, due to their structural complexity, their dynamics cannot be described by a small, finite number of relaxation times. In these systems, the mechanical dissipation must take into account multi-scale dynamical processes, so that their response to an external mechanical stress involves a broad and dense distribution of relaxation times. Soft glassy materials are systems dominated by structural disorder, metastability and rearrangements, and a general description of their mechanical properties has been developed [47, 48]. This model might appear as a good candidate to describe the cell medium. Indeed, it is an out-of-equilibrium and disordered system, in which the rearrangements (through dynamical crosslinking, actin polymerization and molecular motors activity) are made possible by an external supply of chemical energy.

However, at the present stage, the analogy between the general description of soft glassy materials and the cytoskeleton network dynamics remains quite formal, since the elementary biophysical and biochemical mechanisms which govern the cytoskeleton rearrangements are not explained in this description. Some authors have developped a more phenomelogical approach, where the cytoskeletal network is seen as a polarized liquid crystal, and its dynamics is coupled to the activity of molecular motors [49]. Quantitative predictions about the microrheological behavior of the cellular medium has yet to be obtained from this model.

Here we propose another description, intermediate between formal comportemental approaches and more phenomenological structural models. We consider that the cytoskeleton is made of many interconnected units of different length scales (from actin individual filaments to actin bundles and stress fibers). Their size continuously spread from the nanometer scale to the scale of the whole cell. We describe the mechanical response of each unit, labelled by the index *i*, by a simple Kelvin-Voigt model with a response time $\tau_i$. Given the cytoskeleton structure, it is reasonnable to assume that the characteristic response times $\tau_i$ are widely and densely distributed. The elementary creep function $j_i(t)$ associated to each unit $i$ is such that $\frac{dj_i}{dt} = exp(-\frac{t}{\tau_i})$

The description in terms of Kelvin-Voigt units may appear oversimplified, since it does not seem to take into account the active molecular mechanisms related to the molecular motors activity or to the filaments remodelling. However, a precise description of the elastic and dissipation processes at the molecular level may not be necessary to derive macroscopic mechanical behaviors: the spring and the dashpot associated in each unit schematically represent the storage of the elastic energy in actin filaments, bundles, and stress fibers, and its dissipation. This dissipation may include several processes, like cytoskeleton remodelling, molecular motors activity and passive viscosity.

The choice of a generalized Kelvin-Voigt model, where the units are associated in series (figure 10a), is based on convenience, and is adapted to describe creep experiments. The choice of the dual representation, namely the generalized Wiechert-Maxwell model (in which Maxwell elements are placed in parallel) would have yielded the same results. Those two representations are equivalent, since any given model can be reduced to an equivalent series or parallel model [55]. It should be clear that it is not because we consider visco-elastic elements in series that the complex filament network of the cytoplasm is organized in such a way.

The next important ingredient of this model is the distribution $P(\tau_i)$ of relaxation times $\tau_i$. We assume in the following that the cytoskeletal structure in the cell is close to a self-similar one. This assumption is especially supported by fluorescent images of the actin cytoskeleton, showing similarities between the large stress fibers structures at the scale of the cell and the structure of individual filaments at the nanometer scale. This implies that the elementary units introduced above are distributed according to a power law: the number of units having a given size $l$ is taken proportional to $l^{-\xi}$, where $\xi$>0 represents the fractal dimension of the network. Concerning the dependance of the response time $\tau_i$ with the size $l_i$ of elementary units, it is reasonnable to assume that it is depicted by a simple power law: $\tau_i \propto l_i^{\beta}$. As a consequence the distribution of times $\tau_i$ in the cell will itself be a power law of $\tau_i$ : $P(\tau_i) \propto \tau_i^{\alpha-2}$, with $\alpha = 1 - \xi/\beta$. Actually, such a power law distribution is a commonly used assumption in several models of complex viscoelastic solids [50]. This multi-scale coupling between elasticity and dissipation processes is a main characteristic of structural damping.

Assuming first that the relaxation times are continuously distributed according to a power law $P(\tau) = B\tau^{\alpha-2}$, it is straightforward to show that the creep function response *J(t)* of the system also follows a power law of time. Indeed the resulting time derivative of the creep function may be calculated as:

$$\dot{J}(t) = \frac{dJ(t)}{dt} = \int P(\tau)exp(-\frac{t}{\tau})d\tau = B\Gamma(\alpha-1)t^{\alpha-1} \qquad (5)$$

where $\Gamma$ represents the Euler function (defined in the appendix). By integrating this relation over the time *t*, one finds:

$$J(t) = \frac{B\Gamma(\alpha-1)}{\alpha}t^{\alpha} = A_0(\frac{t}{t_0})^{\alpha} \qquad (6)$$

According to Equation (3), the viscoelastic complex modulus $G_e(\omega)$ may then be expressed as a power law of frequency : $|G_e(\omega)| = G_0(\frac{\omega}{\omega_0})^{\alpha}$

In our model, the exponent $\alpha$ of the power law is related to the fractal dimension $\xi$ of the network and to the exponent $\beta$ characterizing the dependance of the response time $\tau$ with the scale *l*. Lacking more information about $\xi$ and $\beta$, it is not possible at this stage to make a quantitative prediction for $\alpha$.

Now we turn to the case of a discrete distribution of relaxation times $\tau_i$. The calculations will be developped in two steps:

- in a first step (i) we assume an ideal power law repartition $P_0$ of the response time $\tau_i$, and we show indeed that the resulting creep function $J_0(t)$ is a power law of time.

- in a second one (ii) we use a set of distributions $P_k$, extracted from $P_0$ by randomly deleting a given fraction of the response times $\tau_i$. We calculate for each distribution $P_k$ the new creep function $J_k(t)$, and analyse the distribution of $J_k(t)$ over different $P_k$. As will be shown in the discussion, this distribution mimicks the distribution of experimental *J(t)* measured for a set of cells belonging to the same cell type.

i) Let us first assume that the relaxation times $\tau_i$ of elementary units are exactly distributed along the time axis according to $\tau_i = \tau_m \, i^{\left(\frac{-1}{1-\alpha}\right)}$, with $0<\alpha<1$. The label *i* varies from 1 to $\infty$, so that $\tau_m$ represents the largest relaxation time in the system. Figure 10b shows a schematic drawing of this distribution $P_0$. The particular form of this repartition has been chosen to reduce in the continuum limit to a power law distribution $P(\tau) = \frac{di}{d\tau} \propto \tau^{\alpha-2}$. In this limit, one recovers an exact power law for the corresponding creep function $J_0(t)$, as obtained by integrating:

$$\dot{J}_0(t) = \sum_i \frac{dj_i}{dt} \cong \int_1^{\infty} exp(-\frac{t}{\tau_i})di = \int_0^{\tau_m} (1-\alpha)(\frac{\tau}{\tau_m})^{\alpha-2} exp(-\frac{t}{\tau})\frac{d\tau}{\tau_m} = B_0 t^{\alpha-1} \qquad (7)$$

To take into account the actual discrete character of the $\tau_i$ distribution, we have performed numerical simulations, where we calculated the exact value of $\dot{J}_0(\theta)$, as a function of the reduced time $\theta = t/\tau_m$. The resulting function, shown in figure 11 (top curve) for a typical value $\alpha = 0.20$ of the exponent, was obtained by summing $10^5$ elementary units ($i$ = 1 to $10^5$). This covers six orders of magnitude (from 1 to $10^{-6}$) for the reduced response times $\tau_i/\tau_m$. As expected, $\dot{J}_0(\theta)$ is perfectly adjusted by a power law of exponent $\alpha$ -1 = - 0.8, in the range $10^{-6} < \theta < 1$. An increased number of elementary units would only extend the range of validity of the power law at the smallest times : it is therefore unnecessary.

ii) In order to build a more realistic picture of the cytoskeleton dynamics, we assume now that only a proportion $p$ of the elementary Kelvin-Voigt units are actually present in a given cell. Indeed, the response times $\tau_i$ are very unlikely to follow the smooth distribution $P_0$ in a real cell. One has to take into account the dispersion of results actually observed from one cell to the other. A given cell should then be represented by its actual distribution $P_k$ of time constants, constructed by selecting, with a given probability $p$, a random set of relaxation times $\tau_i$ from the distribution $P_0$. An example of $P_k$ distribution is schematically represented in figure 10c, together with the original $P_0$ distribution. Under these assumptions, the creep function $J_k(t)$ of the $k$-cell will be calculated from $\dot{J}_k(t) = \sum_i p_i \frac{dj_i}{dt}$, where $p_i$ is a random variable equal to 1 with a probability $p$ and to 0 with a probability 1-$p$. The underlying response times $\tau_i$ remain distributed according to $\tau_i = \tau_m\, i^{\left(\frac{-1}{1-\alpha}\right)}$.

Numerical calculations of $\dot{J}_k(\theta)$ versus the reduced time $\theta = t/\tau_m$ were done for 500 different realizations of $P_k$ distributions ($k$ = 1 to 500), with the same probability $p$ = 0.1. A set of 20 of them are shown in figure 11 (lower curves). These curves show that, at least in the range $10^{-6} < \theta < 10^{-2}$, all the $\dot{J}_k(\theta)$ functions roughly behave as power laws of $\theta$, and exhibit approximately the same exponent $\alpha$-1=-0.8 than $\dot{J}_0(\theta)$. Each curve $\dot{J}_k(\theta)$ was fitted by a power law $\dot{J}_k(\theta) = b_k \theta^{\alpha_k - 1}$ in the range $10^{-6} < \theta < 10^{-2}$ , leading to an exponent $\alpha_k$ and a prefactor $b_k$ for each realization $k$. The histograms of $\alpha_k$ and $b_k$ are presented in figure 12a and 12b. The distribution of $\alpha_k$ is symetric, and is well fitted by a gaussian curve centered on the value $\alpha = 0.20$. On the contrary, the histogram of $b_k$ is clearly asymetric, but one recovers the symetry by plotting the histogram of $\ln(b_k)$ (figure 12c). A first prediction of these simulations is that, over several realizations mimicking the natural dispersion over different cells, the exponents $\alpha_k$ and prefactors $b_k$ are respectively normally and log-normally distributed. As shown in section VI, this is quite consistent with the normal and log-normal distributions of $\alpha$ and $G_0$ experimentally measured on a given set of cells. The width of the distributions calculated in the simulations depends on the drawing probability $p$. In the following, the particular value $p = 0.1$ has been chosen to match the standard deviations of experimental data.

Another important feature emerging from these simulations is that the exponent $\alpha_k$ and prefactor $b_k$ of the power law $\dot{J}_k(\theta)$ are not independent parameters. Figure 13 represents a plot of $ln(b_k)$ *vs* $\alpha_k$ for 500 different realisations ($k$ = 1 to 500): $ln(b_k)$ and $\alpha_k$ appear strongly correlated through a linear relationship. The slope $s = \frac{d(ln(b_k))}{d\alpha_k}$ is found to be equal to 9.8 for the choice of drawing probability $p$ = 0.1. Other numerical tests (not shown here), indicate that this slope $s$ is almost independent of the choice of $p$ in a wide range: $0.01<p<0.8$. We will use $s = 10 \pm 0.5$ to compare this result with experiments in the next section.

## VI - Discussion

We have noticed in section III that, for a given set of cells belonging to the same cell type and submitted to the same experimental protocol, the measured exponents $\alpha$ and prefactors $G_0$ at $f$=1Hz respectively follow a normal and a log-normal distribution (figures 5, 7, 8 and 9). In the model presented in section V, the variability from one cell to another is simulated by the choice of a particular distribution $P_k$ of response times $\tau_i$, randomly selected amongst a general power law distribution $P_0$. We have already pointed out that this model also leads to normal and log-normal distributions for the exponents $\alpha_k$ and prefactors $b_k$ determining the creep function $J_k(t)$. Introducing a dimensionless time $\theta = t/\tau_m$, normalized by the largest time relaxation $\tau_m$ of the cell, we have shown that $\frac{dJ_k}{dt} = \dot{J}_k(\theta) = b_k \theta^{\alpha_k - 1}$, and consequently $J_k(\theta) = \frac{b_k}{\alpha_k}\theta^{\alpha_k}$. Using eq. A6, we predicted the relationship between the exponent $\alpha_k$ and the viscoelastic modulus at $\omega/2\pi = 1$Hz :

$$G_{0k} = \left|G_e(\omega = 2\pi)\right| = \frac{\alpha_k \tau_m^{\alpha_k - 1}(2\pi)^{\alpha_k}}{\Gamma(1+\alpha_k)} \frac{1}{b_k} \qquad (8)$$

As far as the exponents $\alpha_k$ remain close to the averaged value $\alpha$, a log-normal distribution for $b_k$ corresponds to a log-normal distribution for $G_{0k} \propto 1/\, b_k$. This demonstrates that so far the model predictions are consistent with experimental data.

To step further into the comparison beween the model and the data, it is noteworthy to focus on the correlations between exponents $\alpha_k$ and prefactors $G_{0k}$ . Figure 14 gathers the experimental data of $ln(G_0)$ versus $\alpha$ for all the C2 cells, as determined either in optical tweezers or uniaxial stretching experiments. Despite a noticeable dispersion of the results, $ln(G_0)$ appears to be an increasing function of the exponent $\alpha$. This is consistent with the model presented in section V, which predicts a correlation between $ln(b_k)$ and $\alpha_k$. Imposing a linear relationship between $ln(G_0)$ and $\alpha$ in Figure 14, one measures a slope $s' = \frac{\mathrm{d}(\ln(\mathrm{G}_{0k}))}{d\alpha_k}$ = 5.2. A quantitative comparison with the model is then possible, since $s$ and $s'$ are related through :

$$s' = \left.\frac{\mathrm{d}(\ln(\mathrm{G}_{0k}))}{d\alpha_k}\right|_{\alpha_k = <\alpha>} = -s + ln(\tau_m) + ln(2\pi) + \frac{1}{<\alpha>} - \psi(\alpha + 1) \qquad (9)$$

where $\psi(\alpha) = \frac{d(ln(\Gamma(\alpha))}{d\alpha}$ is the digamma function [51]. It is possible to make the measured value $s'$ = 5.2 consistent with the predicted value $s$ = 10 by adjusting the only unknown parameter in our model $\tau_m$, which represents the longest relaxation time in the cell. This adjustment leads to $\tau_m$ ~ 3200 s for the C2 cell type. Since $\theta = t/\tau_m$, the reduced time range $10^{-6} < \theta < 10^{-2}$ then corresponds to a real time range $0.003 < t < 30$s, which exactly matches the experimental range of our measurements. This reinforces the validity of our approach.

The same analysis, performed on our data on alveolar epithelial cells A549, also showed a correlation between the measured exponents $\alpha_k$ and prefactors $G_{0k}$ (data not shown). From the slope $s' = \frac{d(ln(G_{0k}))}{d\alpha_k}$ = 2.8 we infer that the highest response time is $\tau_m$ ~ 300 s for this type of cells. Other experiments [23] also depicted a linear relationship between the viscoelastic modulus $G_1$ (measured at 1 kHz) and the exponent $\alpha$ for HASM cells. In their case, the slope $s''=\frac{d(ln(G_{1k}))}{d\alpha_k}$ ~ 12, measured at 1kHz (figure 10 in [23]), yields a typical time $\tau_m$ ~ 3200 s.

It is remarkable that these estimates of the longest time response in different cell types, derived from different experiments, are roughly consistent with each other and lie in the range 5 min-1hour. Moreover, we emphasize that their common order of magnitude is quite reasonnable, as far as it effectively corresponds to a typical relaxation time at the scale of the whole cell. Indeed, an independant rough evaluation of $\tau_m$ may be obtained by dividing a typical value of the cytoplasm viscosity at long time scale (~ 10 kPa.s) [3, 9] by a typical Young modulus (~$10^2$ Pa) measured at the cell scale in quasi-static experiments. Beyond this time range, some macroscopic remodelling processes (treadmilling, signalization cascades) are known to take place and to interfere with the cell mechanical properties.

## VII – Summary and conclusions

This study makes a parallel analysis of the results obtained by two different microrheological experiments on single living cells, and compares them with other data gathered from various works in the litterature. It allows to bring out some striking common features in the mechanical properties of the cytoskeleton network.

The mechanical response function presents a quasi-universal power law behavior, whatever the experimental technique and the cellular type: the viscoelastic complex modulus of the cell is a power law function of the exciting mechanical frequency f. Correlatively the creep function of the intracellular medium is a power law of elapsed time, with the same exponent $\alpha$. The value of this exponent $\alpha \approx 0.2$ is remarkably homogeneous throughout the large panel of studied cell types, whatever the experimental techniques and conditions. It does not seem to either depend on the probe size, nor on the typical length scale on which the experiment is performed. This clearly demonstrates that there is no characteristic dissipation time in the cellular response, or more precisely that these relaxation times are broadly distributed over a wide time interval, extending at least from 0.01s to 100s. This is a characteristic of structural damping, where the mechanisms responsible for the storage of elastic energy and its dissipation are strongly correlated.

Moreover, the prefactor of the response function, which represents the value of the elastic modulus at a given reference frequency, varies by almost two orders of magnitude, according to the cellular type and to the experimental conditions. Interestingly, for a set of cells of a given type, probed in the same experiment, the prefactor distribution is found log-normal. The possible influence of several physical or biological parameters on the average value of this prefactor is discussed: temperature, probe size, nature of the mechanical receptor, cell inhomogeneities. Further studies should concentrate on systematic comparisons between different cell types, probed in identical experimental conditions: this is the only way to - perhaps - associate a given cell function with a characteristic viscoelastic behavior.

The semi-phenomenological model presented here is able to accurately predict the mechanical response of a living cell submitted to a controlled stress, in a wide range of time scales. The mechanical behavior of the cell is modelled by associating a large number of elementary Kelvin-Voigt units, which account for the different scales of the cytoskeletal network. Assuming a self-similar structure of the network and a power law distribution of time constants, one recovers all the features of the macroscopic behavior observed on different cell types. This approach quantitatively accounts for the power law responses measured in different rheological experiments, and also for the normal and log-normal distributions retrieved for the exponents and prefactors. The largest relaxation time in the cell, which is the only adjustable parameter of the model, is consistent with other independant estimates. A further step will consist in interpreting the dissipative elements in term of elementary biological mechanisms, such as molecular motors activity and crosslinkers dynamics, which play a crucial role in the cytoskeleton remodelling.

We acknowledge Mireille Lambert for kindly providing C2 cells and cadherin protocols. This work was partly supported by grants from Université Paris 7 (Bonus Qualité Recherche) from the French Ministère de la Recherche (ACI jeune chercheur), from the French Centre National de la Recherche Scientifique (programme Physique et Chimie du Vivant) and from the Association pour la Recherche sur le Cancer (subvention libre n° 3115).

## APPENDIX A: Relation between the creep function *J(t)* and the complex modulus $G_e(\omega)$

When submitting a given material to a varying stress $\sigma(t)$, the induced strain $\varepsilon(t)$ is related to $\sigma(t)$, in the linear regime, by:

$$\varepsilon(t) = J(t)\sigma(0) + \int_0^t J(t-t')\dot{\sigma}(t')dt' \qquad \text{(A1)}$$

where *J(t)* is the creep function (*i.e.* the strain generated by a step stress, normalized by the constant stress value). Defining the Laplace Transform by $LT\left[f(t)\right] = \tilde{f}(s) = \int_0^{+\infty} e^{-st} f(t)dt$, equation A1 leads to[1] :

$$\tilde{\varepsilon}(s) = s\,\tilde{J}(s)\tilde{\sigma}(s) \qquad \text{(A2)}$$

On the other hand, in response to an oscillating stress $\sigma = \sigma(\omega)\exp(j\omega t)$, the induced strain can be written $\varepsilon = \varepsilon(\omega)\exp(j\omega t)$, which allows to define a viscoelastic complex modulus $G_e(\omega)$ as $G_e(\omega) = \frac{\sigma(\omega)}{\varepsilon(\omega)}$. In the limit $\omega \to 0$, $G_e(\omega)$ reduces to the Young modulus *E* of the material.

Since the Fourier Transform $FT\left[f(t)\right] = \hat{f}(\omega) = \int_{-\infty}^{+\infty} e^{-j\omega t} f(t)dt$ is related to the Laplace Transform by $\hat{f}(\omega) = \tilde{f}(j\omega)$, one can rewrite eq. (A2) as $\hat{\varepsilon}(\omega) = j\omega\,\hat{J}(\omega)\hat{\sigma}(\omega)$. Consequently, a general relation exists between the viscoelastic modulus $G_e(\omega)$ and the creep function *J*:

$$G_e(\omega) = \frac{1}{j\omega\,\tilde{J}(j\omega)} = \frac{1}{j\omega\,\hat{J}(\omega)} \qquad \text{(A3)}$$

Now we restrict ourself to the particular case where *J(t)* behaves as a power law of time: $J(t) = A_0(\frac{t}{t_0})^{\alpha}$. Here $t_0$ is an arbitrary reference time, chosen for convenience equal to 1s. The Laplace Transform of *J(t)* is equal to :

$$\tilde{J}(s) = \frac{A_0\Gamma(1+\alpha)}{s(st_0)^{\alpha}} \qquad \text{(A4)}$$

---

[1] According to the usual definition of the compliance *J**, one has $J^*(s) = s\tilde{J}(s)$

where $\Gamma(1+\alpha)=\int_0^{+\infty} e^{-x}x^{\alpha}dx$ is the Gamma Euler function.

In this case the corresponding viscoelastic complex modulus takes the form:

$$G_e(\omega)=|G_e|\,e^{j\delta}=\frac{(j\omega t_0)^{\alpha}}{A_0\Gamma(1+\alpha)} \tag{A5}$$

with a complex norm $|G_e|=\dfrac{\omega^{\alpha}t_0^{\alpha}}{A_0\Gamma(1+\alpha)}$ (A6)

and a phase $\delta=\dfrac{\alpha\pi}{2}$ (A7)

which is independant of the frequency ω.

## APPENDIX B: Force-displacement relationship for a bead partially immersed in an elastic medium.

We recall here the main results of the analytic calculation of the displacement of a rigid spherical bead, immersed in a semi-infinite homogeneous medium, and submitted to a force $\vec{F}$ tangential to the medium boundary (see figure 3) [12]. We assume that the medium is incompressible, so that the static Young modulus *E* and the shear modulus *μ* are related by $E = 3\mu$.

*i) Bead immersed in an infinite medium*

In the simple case of an infinite incompressible medium, the displacement *x* of the center of the bead submitted to a force *F* is given by [52, 53]:

$$F = 2\pi REx = 6\pi R\mu x \tag{B1}$$

Where R represents the bead radius.

*ii) Bead in contact by a small area with a semi-infinite medium*

This case corresponds to a bead weekly immersed in the medium, *i.e.* where the half-angle $\Theta$ of the immersion cone is small. We have shown [12] that in this limit the displacement *x* of the bead center results of the combination of a translation of the bead and a rotation around the region of contact, and that *F* and *x* are related through:

$$F = 2\pi REf(\Theta)x \tag{B2}$$

Where $f(\Theta)$ is a purely geometrical factor approximately given by :

$$f(\Theta)=\frac{1}{\left(\frac{9}{4\sin\Theta}+\frac{3\cos\Theta}{2\sin^3\Theta}\right)} \qquad \text{(B3)}$$

In the denominator, the first term stands for the contribution of the bead translation, the second one for its rotation. This analytic expression is consistent with other numerical studies describing the pure rotation of a magnetic bead submitted to a torque [31, 54]. In particular, all the works agree on the fact that the bead rotation is proportional to $\sin^3\Theta$. Only the numerical prefactor differs by approximatively a factor of two between the analytical and numerical approaches. We have shown in [12] that equation (B3) can be extrapolated to the full range of angle $\Theta$ accessible to experiment, typically 20° < $\Theta$ < 70°. As a consequence, we use this equation to interpret our data in the main course of this paper.

Equation (B1)-(B3) describe the force-displacement relationship in the static case where $F$ (and $x$) are kept constant with time. When the bead is submitted to an oscillatory force $F = \hat{F}(\omega)\exp(j\omega t)$, they can be generalized to define the viscoelastic complex modulus $G_e$ as :

$$\hat{F}(\omega)=2\pi RG_e(\omega)\hat{x}(\omega) \qquad \text{(B4)}$$

for the case of a bead totally immersed in the medium,

and
$$\hat{F}(\omega)=2\pi RG_e(\omega)f(\Theta)\hat{x}(\omega) \qquad \text{(B5)}$$

for the cas of partial immersion. Here we assume that $f(\Theta)$ keeps the same expression as in (B3). In this definition, $G_e$ reduces to the Young modulus $E$ as the frequency goes to zero. Notice that this convention is different from the one proposed by other authors [31] who define a viscoelastic modulus $G_s = G_e/3$ from the shear modulus $\mu$. For instance, in the case of magnetic beads, the bead rotation $\Phi$ is related to the applied specific torque $T_s$ (torque per unit volume) by :

$$T_S = 6\beta G_S\Phi \qquad \text{(B6)}$$

where $\beta$ is a numerically computed geometrical factor. In the limit of a weekly immersed bead, eq. (B6) takes a form equivalent to eq. (B5) :

$$F = 8\pi RG_S\beta x \qquad \text{(B7)}$$

**TABLE 1**

| Cell type | $\langle \alpha \rangle$ | $\sigma_\alpha$ (std deviation) | $\Delta_\alpha$ (std error) | $\langle \ln G_0 \rangle$ | $\sigma_G$ (std deviation) | $\Delta_G$ (std error) | $G_{0M}$ (Pa) (median value) | $\langle G_0 \rangle$ (Pa) (mean value) | # cells | Technique/coating |
|---|---|---|---|---|---|---|---|---|---|---|
| Myoblasts C2 | 0.242 | 0.082 | 0.013 | 6.46 | 0.82 | 0.12 | 640 (+80/-70) | 1060 ± 190 | 43 | Uniaxial stretching/ Glutaraldehyd |
| Myoblasts C2 | 0.29 | 0.07 | 0.02 | 6.75 | 0.26 | 0.08 | 850 (+75/-65) | | 12 | Uniaxial stretching/ Cadherin |
| Myoblasts C2-7 | 0.208 | 0.098 | 0.021 | 5.73 | 1.68 | 0.36 | 310 (+130/-100) | 570 ± 150 | 22 | Optical tweezers/ RGD |
| Alveolar epithelials A549 | 0.219 | 0.067 | 0.014 | 6.04 | 0.82 | 0.17 | 420 (+80/-70) | 705 ± 155 | 23 | Optical tweezers/ RGD |
| Alveolar epithelials A549 | 0.181 | 0.06 | 0.014 | 4.41 | 1.43 | 0.33 | 80 (+25/-20) | 160± 40 | 19 | Optical tweezers/ Anti-ICAM-1 |
| Macrophages | 0.25 | 0.11 | 0.04 | 7.70 | 0.46 | 0.16 | 2210 (+390/-330) | | 8 | Uniaxial stretching/ Glutaraldehyd |
| Macrophages | 0.20 | 0.08 | 0.03 | 7.55 | 0.73 | 0.27 | 1910 (+600/-460) | | 7 | Uniaxial stretching/ Fibronectin |
| Fibroblasts L929 | 0.15 | 0.06 | 0.02 | 4.16 | 1.14 | 0.36 | 65 (+25/-20) | 170 ± 90 | 10 | Optical tweezers/ RGD |
| Fibroblasts Primary cells | 0.26 | 0.07 | 0.02 | 6.61 | 1.48 | 0.41 | 750 (+380/-250) | | 13 | Uniaxial stretching/ Glutaraldehyd |
| Canin Kidney MDCK | 0.18 | 0.10 | 0.03 | 7.58 | 1.24 | 0.39 | 1950 (+950/-600) | 3660 ± 1150 | 10 | Optical tweezers/ RGD |

**FIGURE CAPTIONS**

"(Color online)"

Figure 1: View of the uniaxial stretching rheometer. A single cell is attached by specific or non specific binding to a rigid glass microplate (bottom) and a flexible one (top). The bar length is 10µm. The stiffness of the flexible plate is calibrated, so that one simultaneously measures the force applied to the cell, and its deformation. A feedback loop allows to apply a constant stress to the cell, and thus to retrieve its creep function $J(t)$.

"(Color online)"

Figure 2: Principle of the measurements of the complex viscoelastic modulus $G(\omega)$, using the optical tweezers setup. A silica microbead, specifically bound to given membrane receptors, is trapped in the focused laser beam, and is used as a handle to apply a force to the cytoskeleton. The experimental chamber is submitted to a sinusoïdal displacement at frequency $f=\omega/2\pi$, while the optical trap is kept at a fixed position. The applied force and the cell deformation are deduced from the displacements of the bead $x_b$ and of the chamber $x_c$.

"(Color online)"

Figure 3: Axial view of a bead (3.47 µm in diameter) partially embedded into the cell edge, and trapped in the optical tweezers. The force $F$ is applied tangentially to the cell membrane. The derivation of the complex viscoelastic modulus $G(\omega)$ takes into account the value of the immersion angle $\Theta$ of the bead into the cell, estimated from the image.

"(Color online)"

Figure 4: Plot of the creep function $J(t)$, measured for a single C2 myoblast with the stretching rheometer. Here the cell is attached to the glass plates through glutaraldehyde coating. Over three decades in time, $J(t)$ is very well fitted by a power law $J(t) = A_0(t/t_0)^{\alpha}$. For this particular cell, the exponent $\alpha$ is found equal to 0.25.

"(Color online)"

Figure 5: Histograms of the distributions of the exponents $\alpha$ (5a) and of the logarithms of the prefactors $ln(G_0)$ (5b), measured with the stretching rheometer on a set of 43 C2 myoblasts, attached to the plates through glutaraldehyde coating. The cumulative distributions of $\alpha$ and $ln(G_0)$ are also plotted (black dots). They are correctly fitted by an error function $1+ erf\ [(x-$

$x_0)/\sqrt{2}\sigma]$ (blue lines). This indicates that the distribution of the exponents $\alpha$ is a normal distribution, and that the distribution of the prefactors $G_0$ is log-normal. The fits give the best estimates for the average exponent $<\alpha>$ = 0.242 ± 0.013, and for the median value of the prefactor $G_{0M} = exp(<ln(G_0)>)$ = 640 (+80/-70) Pa.

"(Color online)"

Figure 6: Plot of the modulus $|G_e|$ and of the phase $\delta$ of the complex viscoelastic coefficient $G_e(\omega)$ measured with optical tweezers, as a function of the frequency $f=\omega/2\pi$, for a single C2 myoblast. The bead is coated with RGD peptides and binds to integrins. $|G_e|$ behaves as a power law of $f$ over three frequency decades, with an exponent $\alpha$ here found equal to 0.30. The phase shift $\delta$ remains constant in the studied frequency range. This is consistent with the power law behavior of the creep function $J(t)$ measured with the stretching rheometer for the same C2 cell type.

"(Color online)"

Figure 7: Histograms of the distributions of the exponents $\alpha$ (7a) and of the logarithms of the prefactors $ln(G_0)$ (7b), measured with the optical tweezers on a set of 22 C2 myoblasts. The beads bind to integrins via RGD coating. The cumulative distributions (black dots) are well adjusted by an error (erf) function (blue line). The best estimate for the average exponent is $<\alpha>$ = 0.208 ± 0.021, and for the median value of the prefactor is $G_{0M}$ = 310 (+130/-100) Pa.

"(Color online)"

Figure 8: Histograms of the distributions of the exponents $\alpha$ (8a) and of the logarithms of the prefactors $ln(G_0)$ (8b), measured with the optical tweezers on a set of 23 epithelial alveolar cells A549. The beads bind to integrins via RGD coating. The distributions of $\alpha$ and $G_0$ are normal and log-normal, respectively. From the erf fits of the cumulated distributions, the best estimate for the average exponent is $<\alpha>$ = 0.219 ± 0.014, and for the median value of the prefactor is $G_{0M}$ = 420 (+80/-70) Pa.

"(Color online)"

Figure 9: Histograms of the distributions of the exponents $\alpha$ (9a) and of the logarithms of the prefactors $ln(G_0)$ (9b), measured with the optical tweezers on a set of 19 epithelial alveolar cells A549. The beads bind to integrins via anti-ICAM1 coating. From the erf fit of the cumulated distributions, the best estimate for the average exponent is $<\alpha> = 0.181 \pm 0.014$, and for the median value of the prefactor is $G_{0M}$ = = 80 (+25/-20) Pa.

"(Color online)"

Figure 10: The cytoskeleton network is modeled as an infinite assembly of elementary units labelled by the index *i*, each of them showing a simple Kelvin-Voigt behavior with a response time $\tau_i$ (Fig 10a). In the ideal case, the relaxation times $\tau_i$ are assumed to be exactly distributed according to a power law $\tau_i = \tau_m \, i^{\left(\frac{-1}{1-\alpha}\right)}$, where $\tau_m$ represents the largest relaxation time in the cell. This distribution $P_0$ is represented in Fig. 10b. To take into account some natural dispersion, a given cell *k* is modelled by a proportion $p<1$ of relaxation times, randomly selected from $P_0$. An example of distribution $P_k$ is represented in Fig. 10c.

"(Color online)"

Figure 11: Plot of the time derivative of the creep function $dJ/d\theta$, numerically calculated for the ideal power law distribution $P_0$ (top curve), and for 20 more realistic distributions $P_k$ (bottom curves) randomly extracted from $P_0$ as explained in the text. The time scale is normalized by the largest time relaxation in the cell $\tau_m$. The simulations involve $10^5$ elementary units, and the reference exponent in the distribution $P_0$ is equal to 0.20. As expected, the response function $dJ_0/d\theta$ corresponding to the distribution $P_0$ exactly merges into a power law (dashed curve) of exponent $\alpha$-1= -0.8, in the range $10^{-6} < \theta < 1$. For the distributions $P_k$ , the response functions $dJ_k/d\theta$ are well adjusted by power laws in the range $10^{-6} < \theta < 10^{-2}$, with a distribution of exponents $\alpha_k$ close to $\alpha$.

"(Color online)"

Figure 12: Histograms of the distributions of the exponents $\alpha_k$ (12a), of the prefactors $b_k$ (12b), and of their logarithms $ln(b_k)$ (12c). These quantities are measured from the numerically calculated curves $dJ_0/d\theta$, for 500 different realizations of $P_k$ distribution. While the distribution of exponents $\alpha_k$ is symmetrical, the distribution of prefactors $b_k$ is clearly asymmetric. The cumulated distributions and their best fits by erf functions are also represented (12a and 12c). They show that, according to this model, $\alpha_k$ and $b_k$ are respectively normally and log-normally distributed.

"(Color online)"

Figure 13: Plot of the prefactors $ln(b_k)$ *vs* exponents $\alpha_k$, for 500 different realizations of $P_k$ distribution. As can be seen, the model predicts a strong linear correlation between these two quantities.

"(Color online)"

Figure 14: Plot of the experimental values of the prefactors $ln(G_0)$ *vs* exponents $\alpha$, measured for C2 myoblasts. The data from optical tweezers experiments (disks) and stretching rheometer experiments (squares) are plotted together. Despite the dispersion of the results, $ln(G_0)$ appears to be an increasing function of the exponent $\alpha$. This is consistent with the prediction of the model.

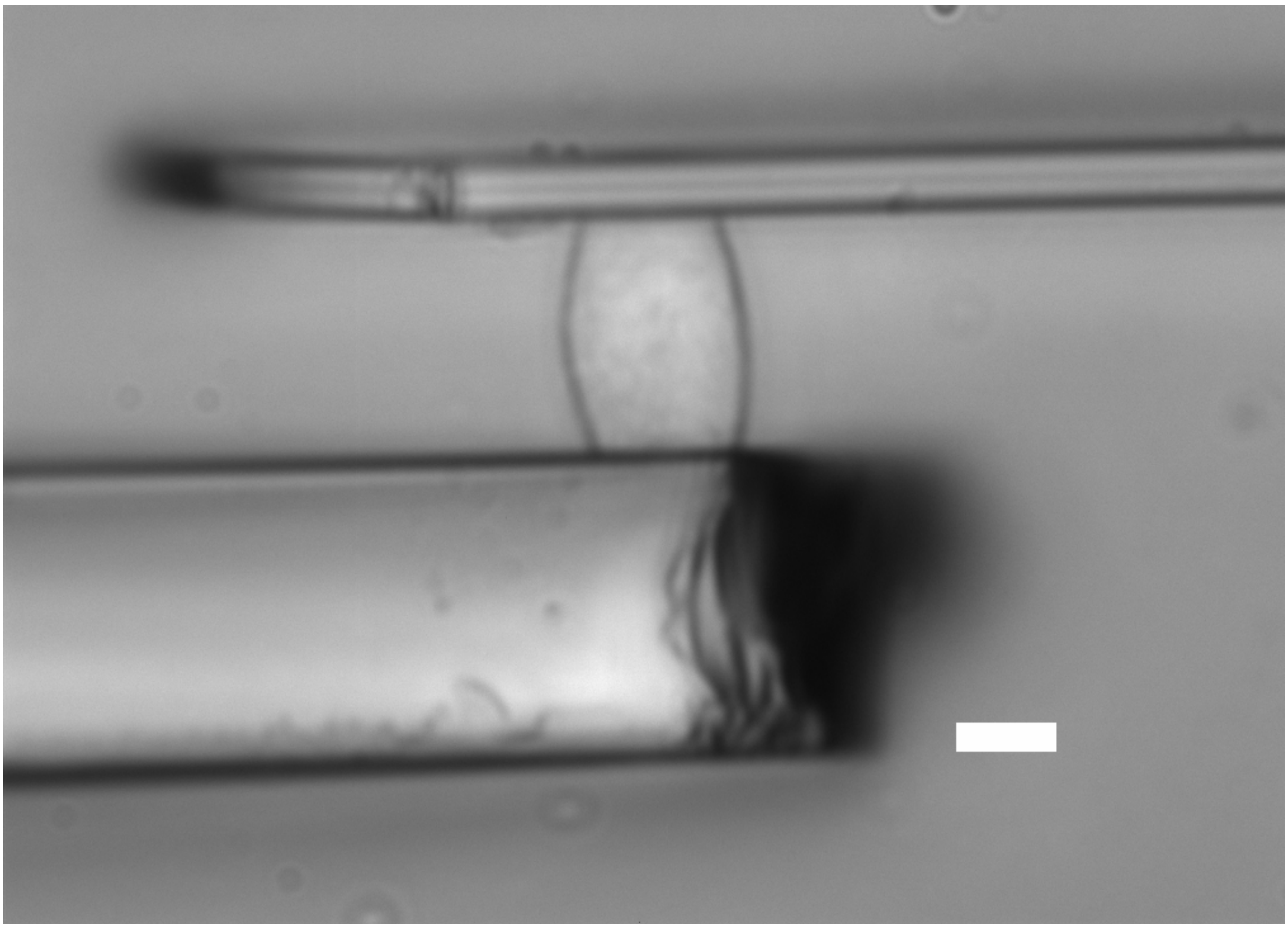

**Figure 1**

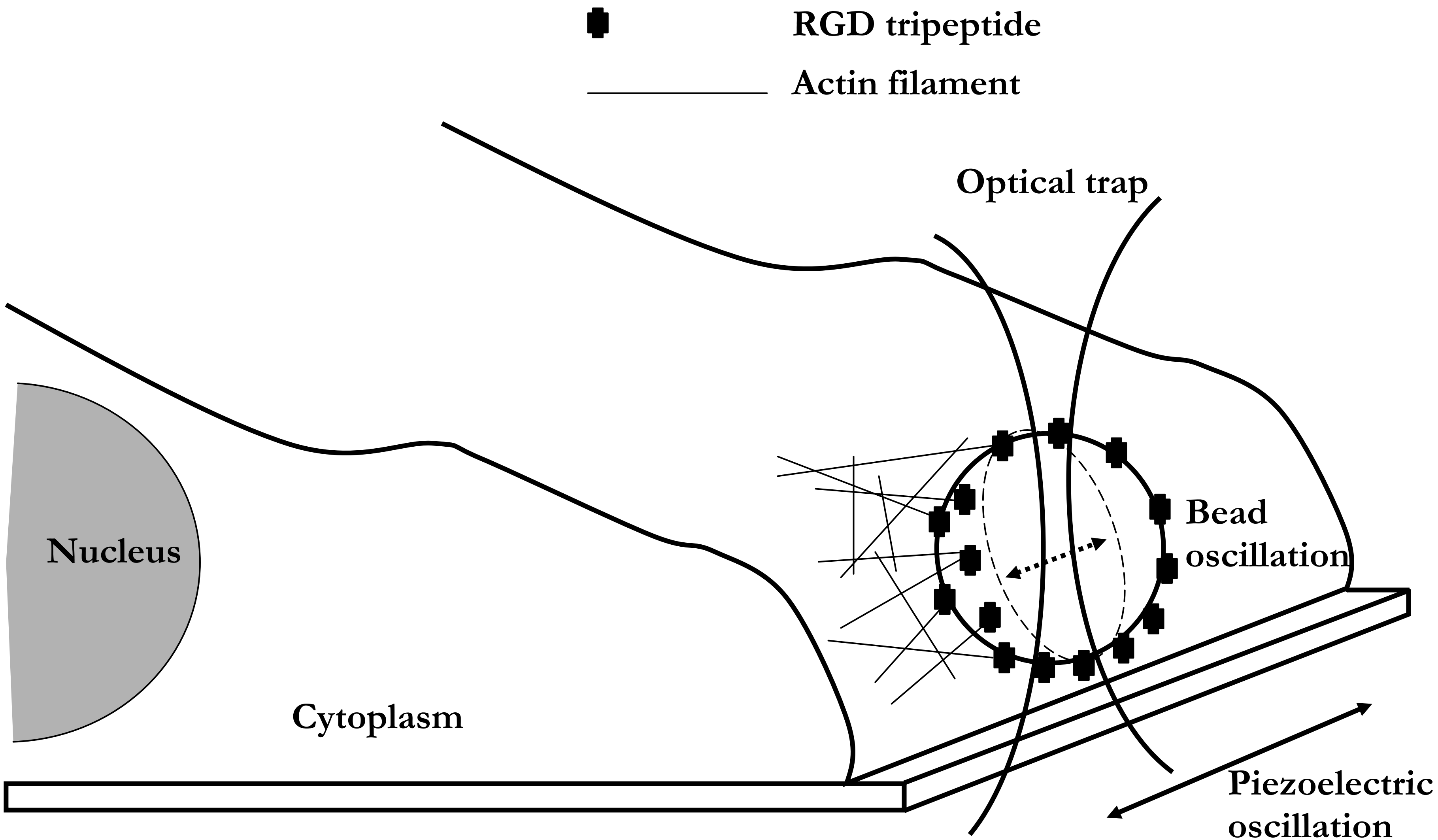

**Figure 2**

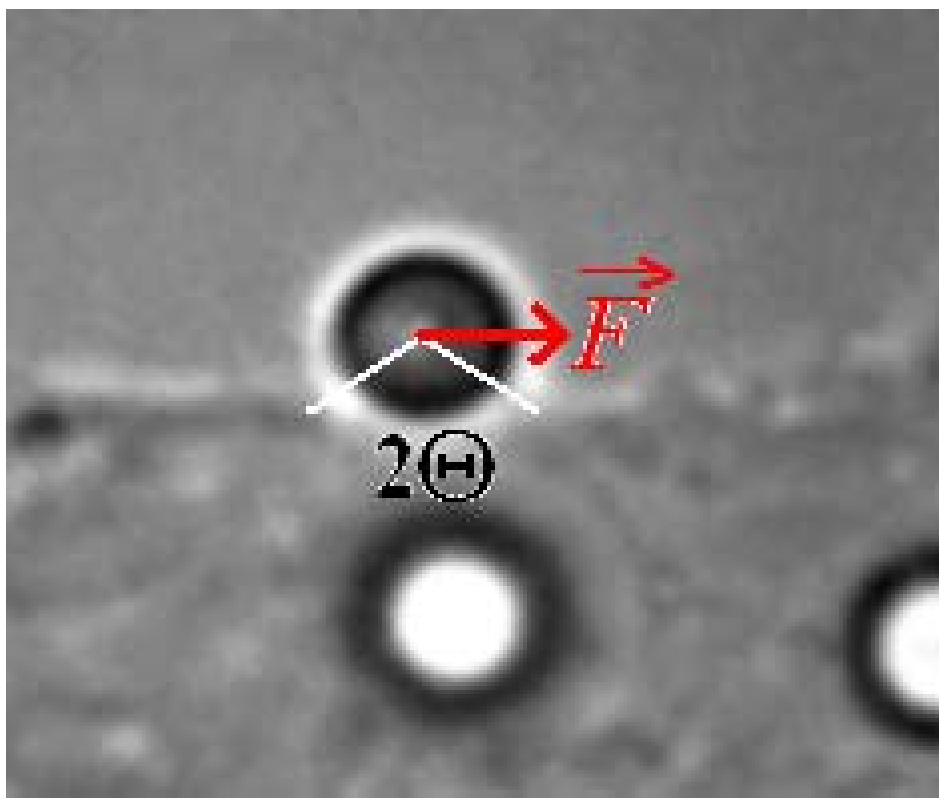

Figure 3

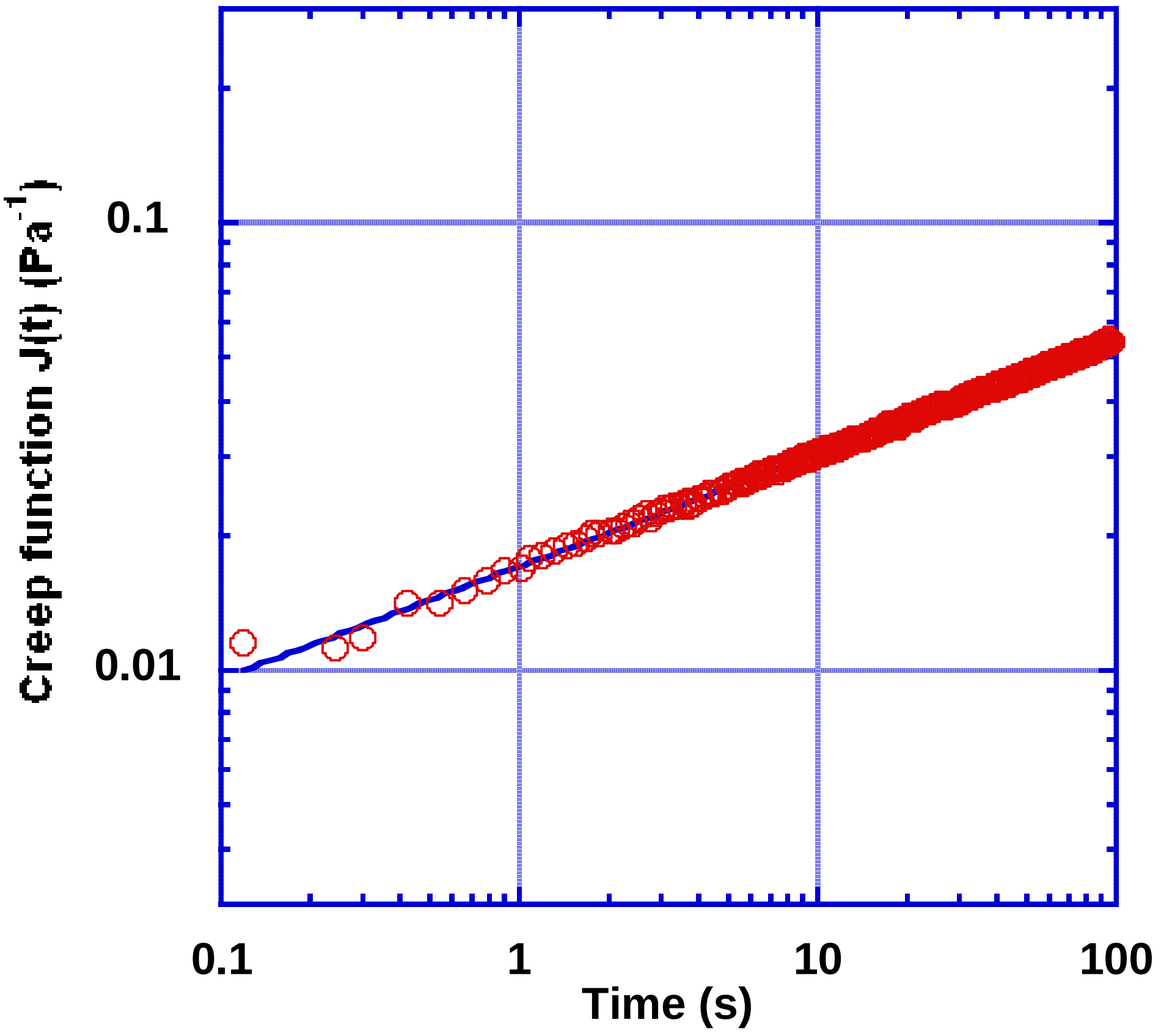

Figure 4

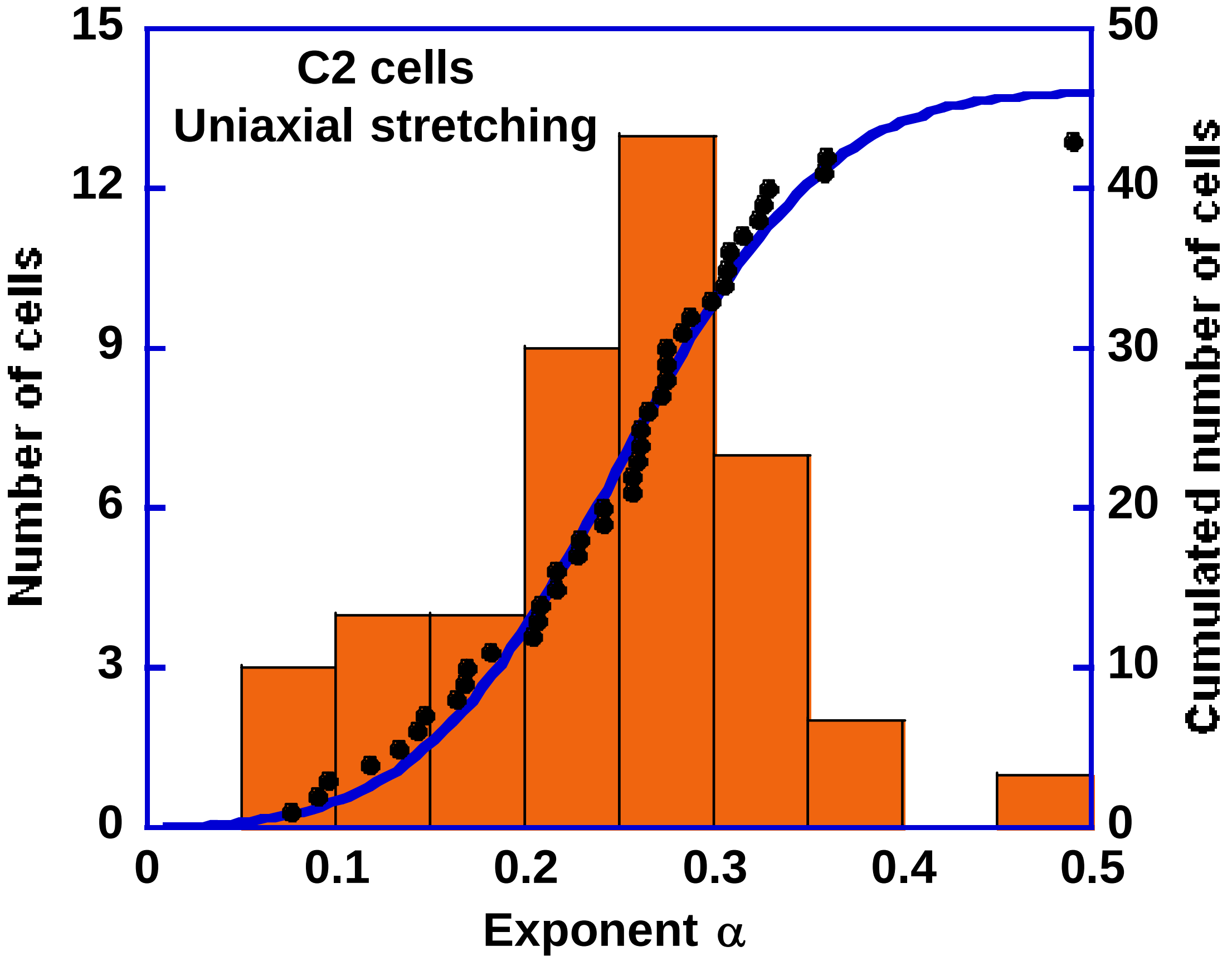

Figure 5a

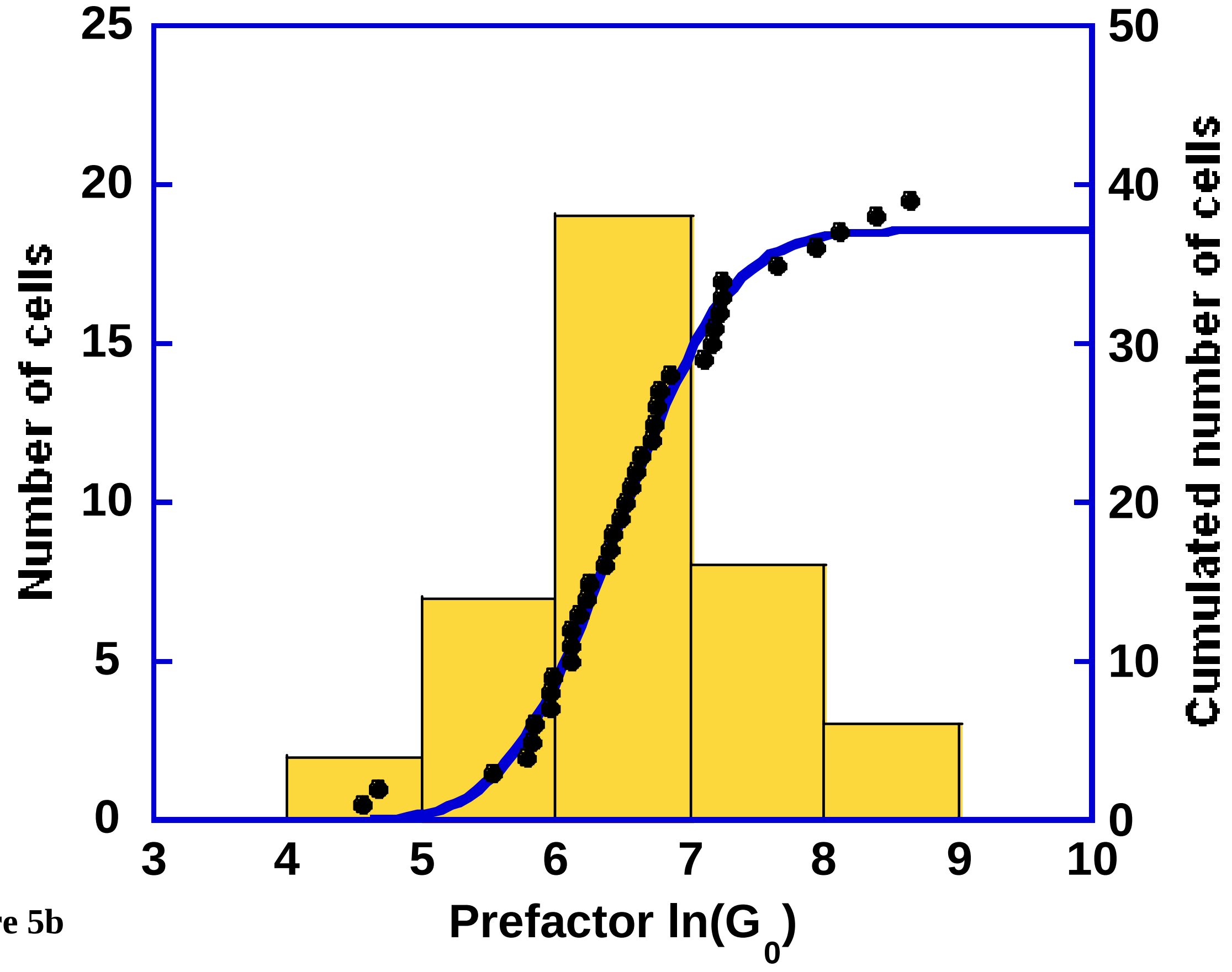

Figure 5b

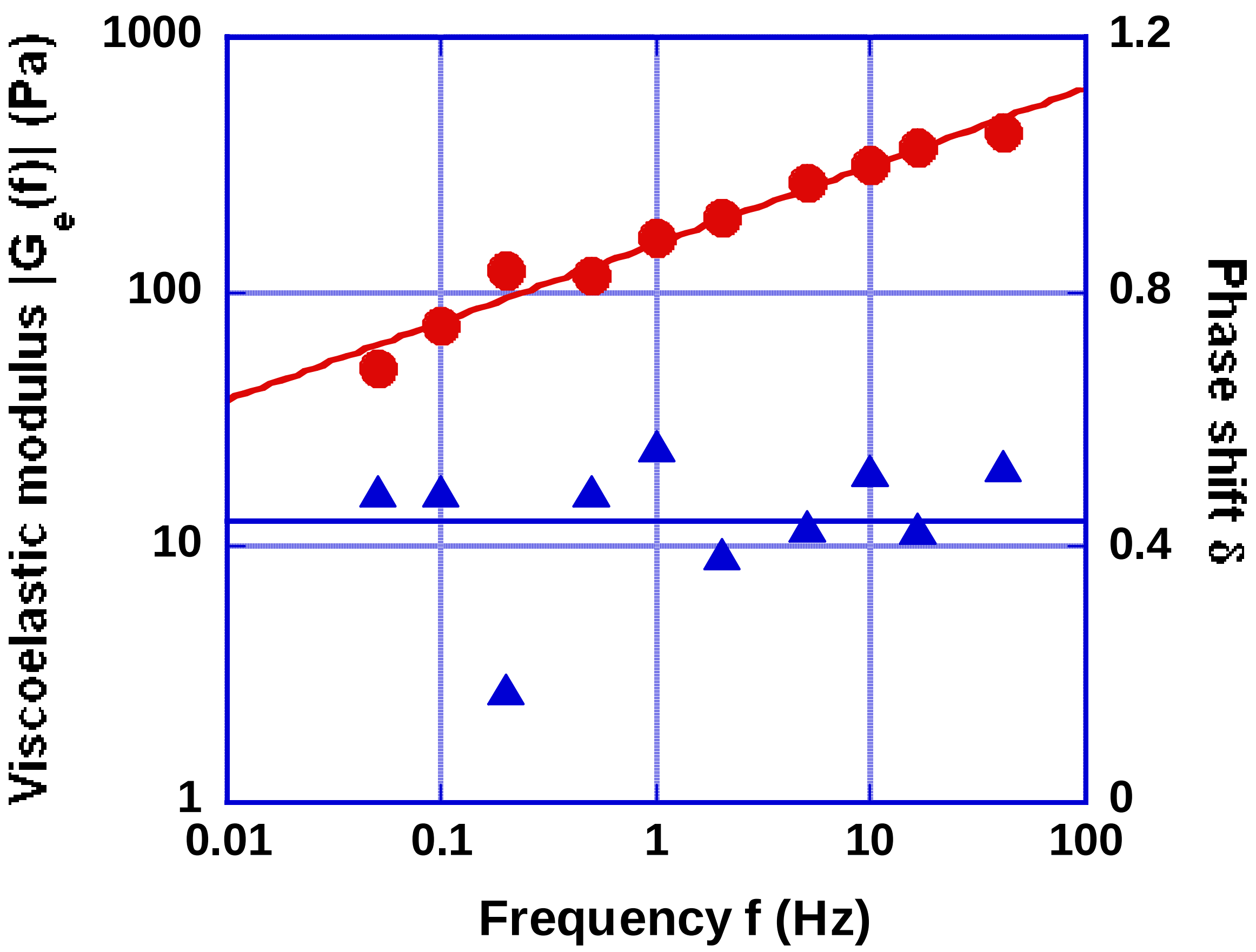

**Figure 6**

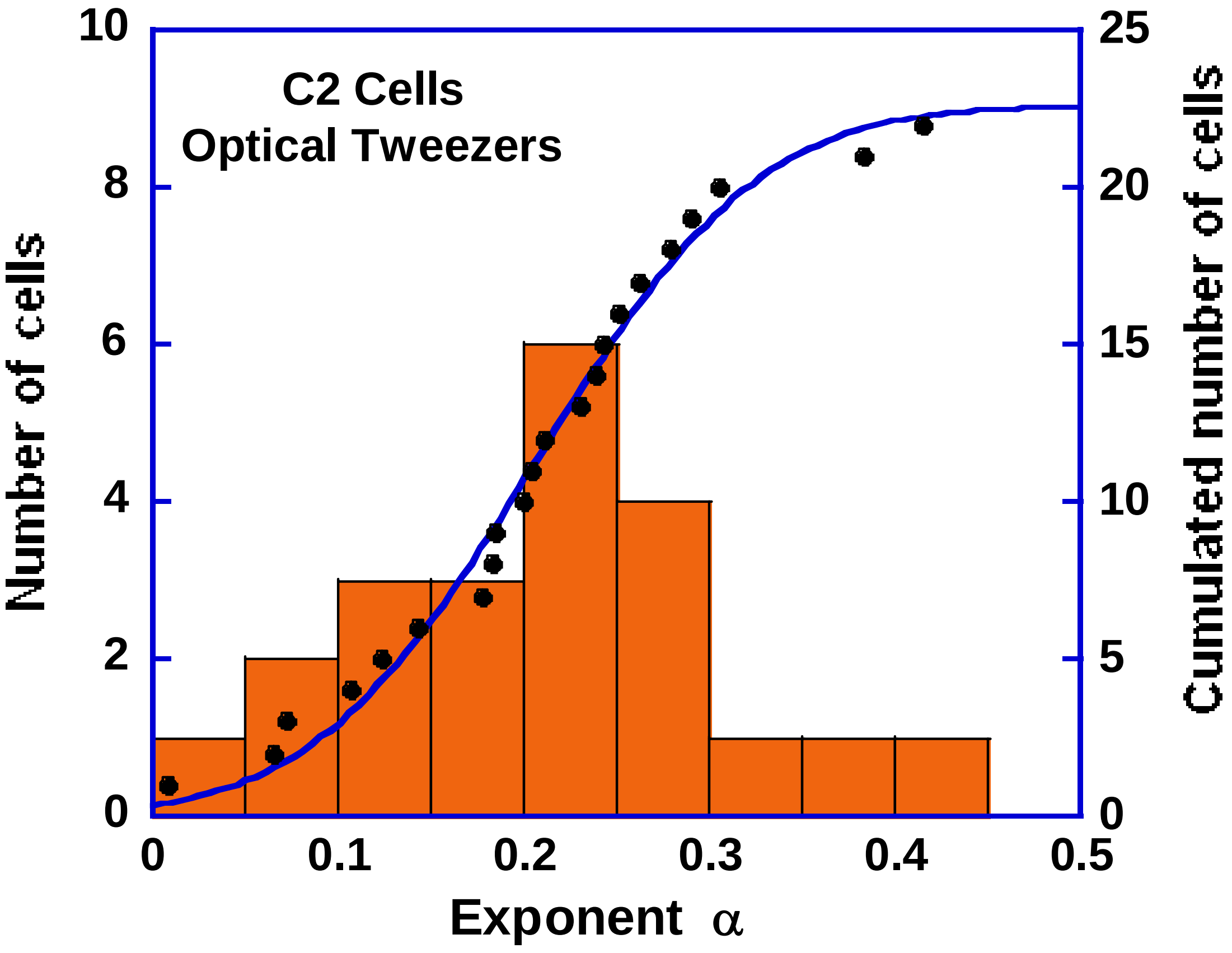

Figure 7a

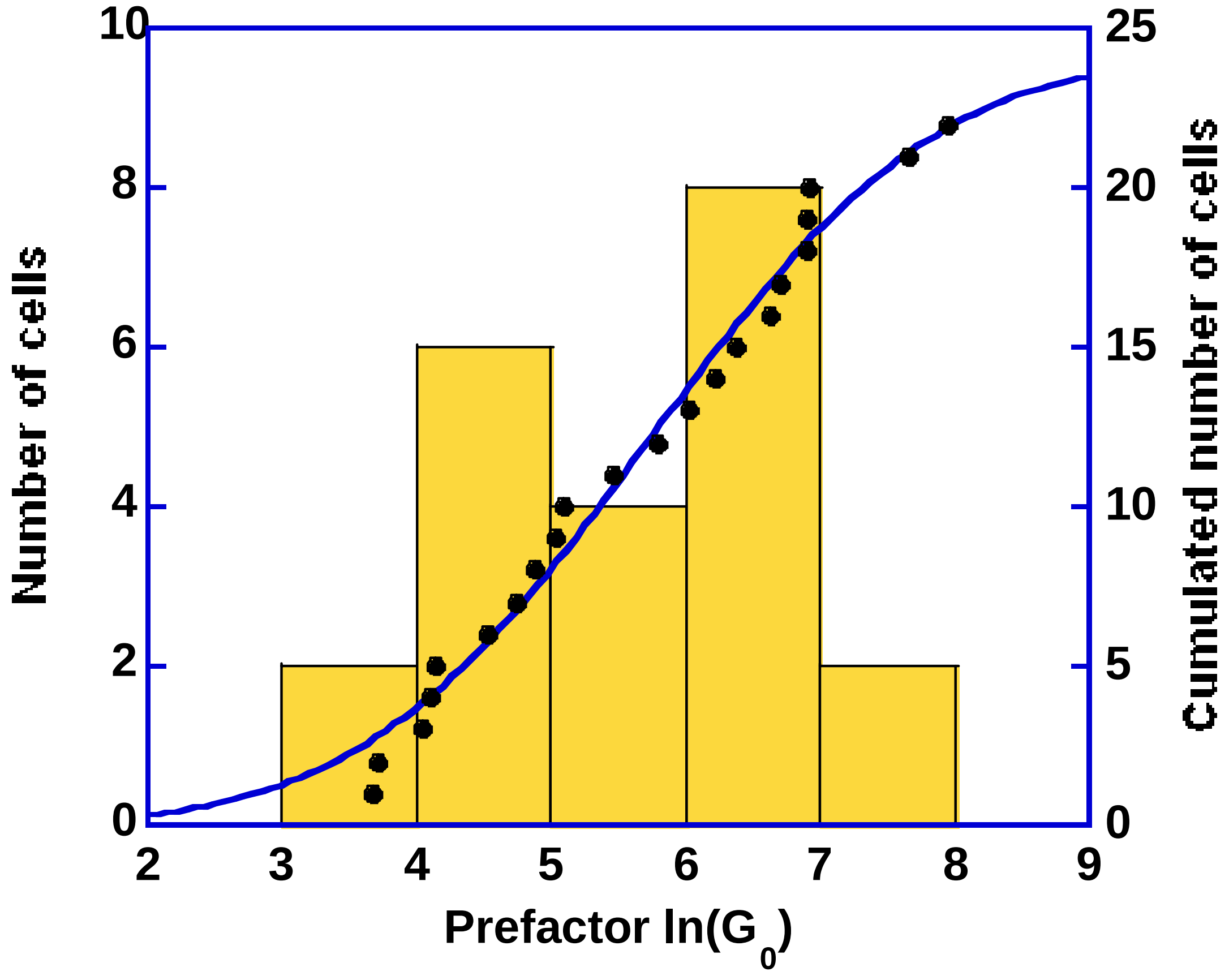

Figure 7b

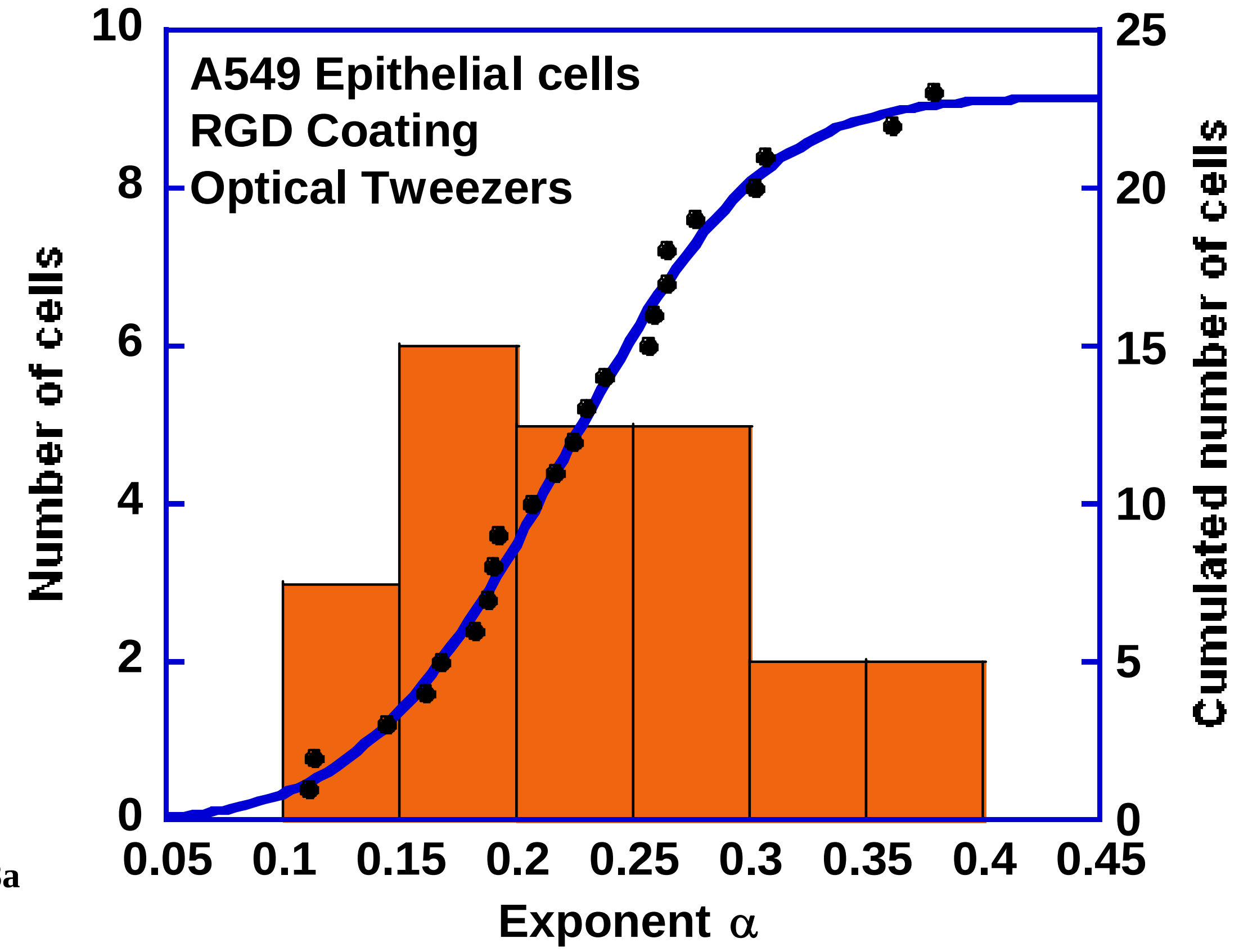

Figure 8a

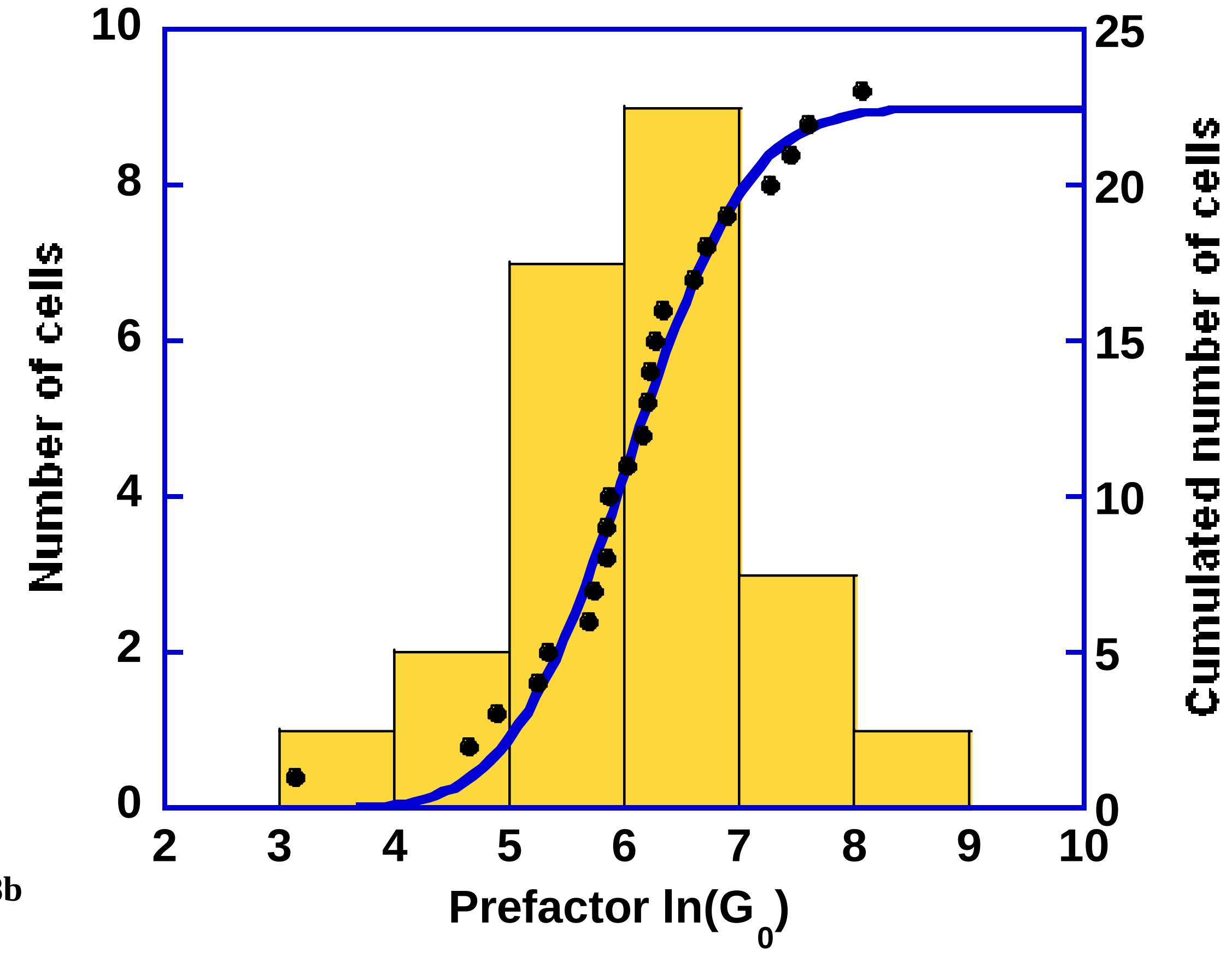

Figure 8b

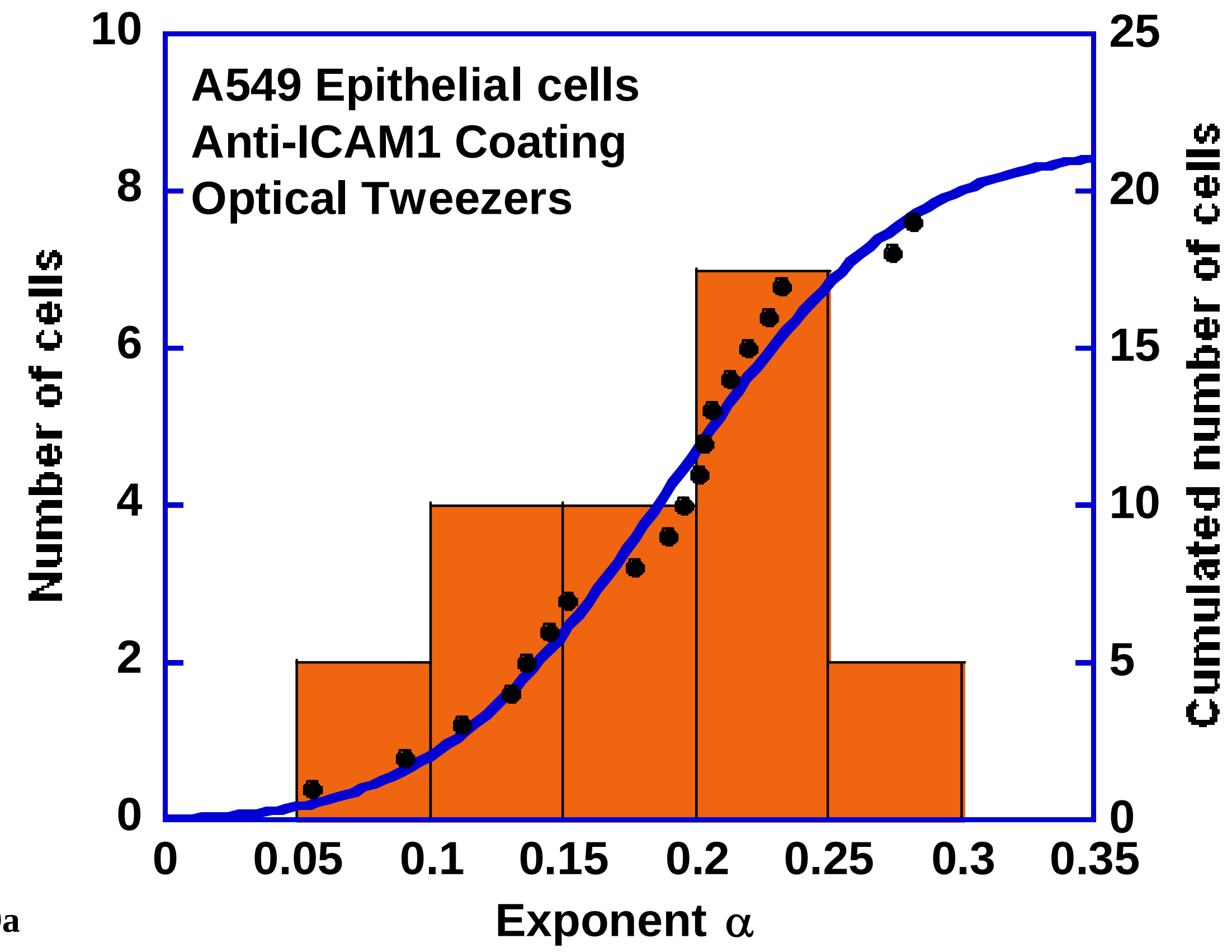

Figure 9a

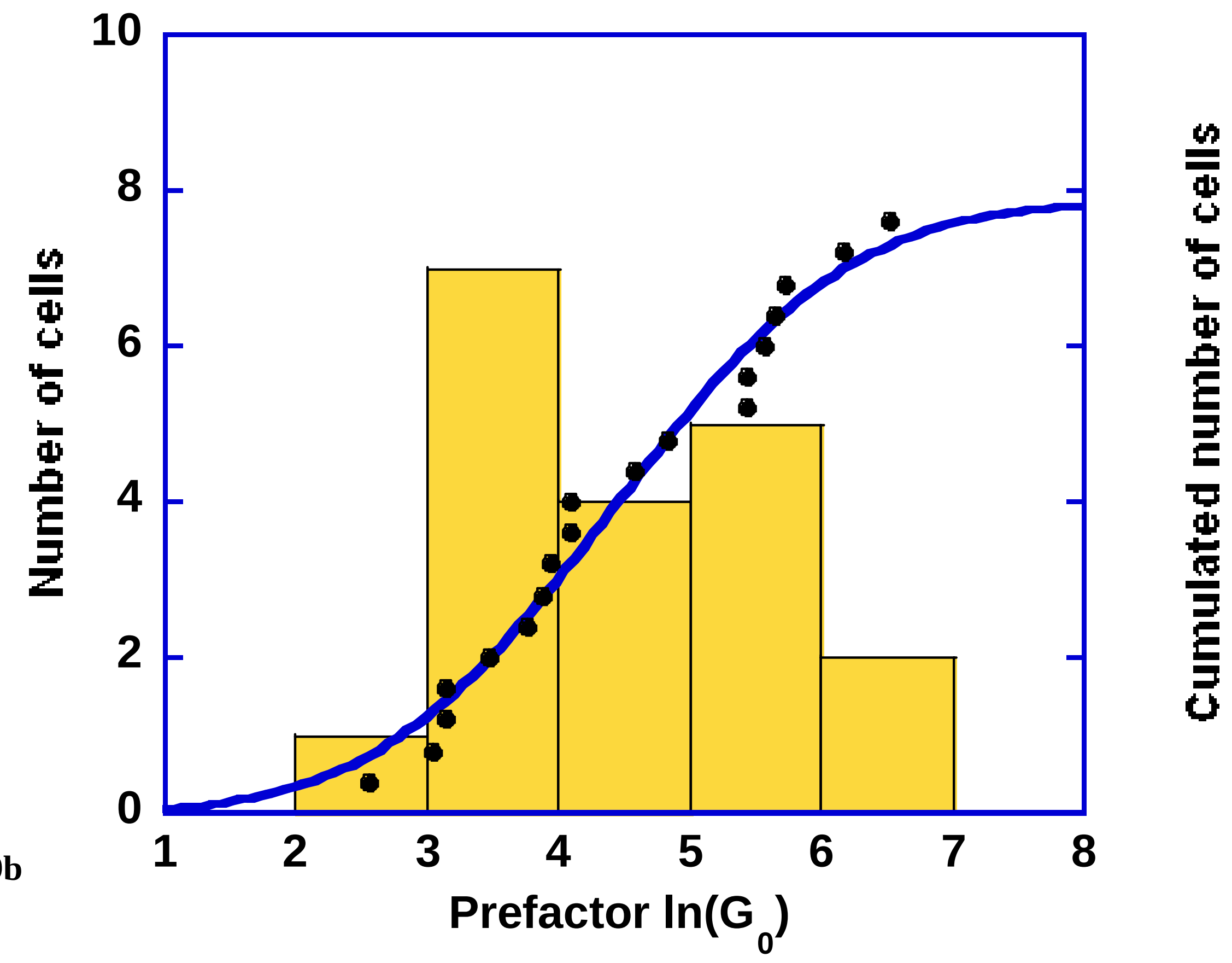

Figure 9b

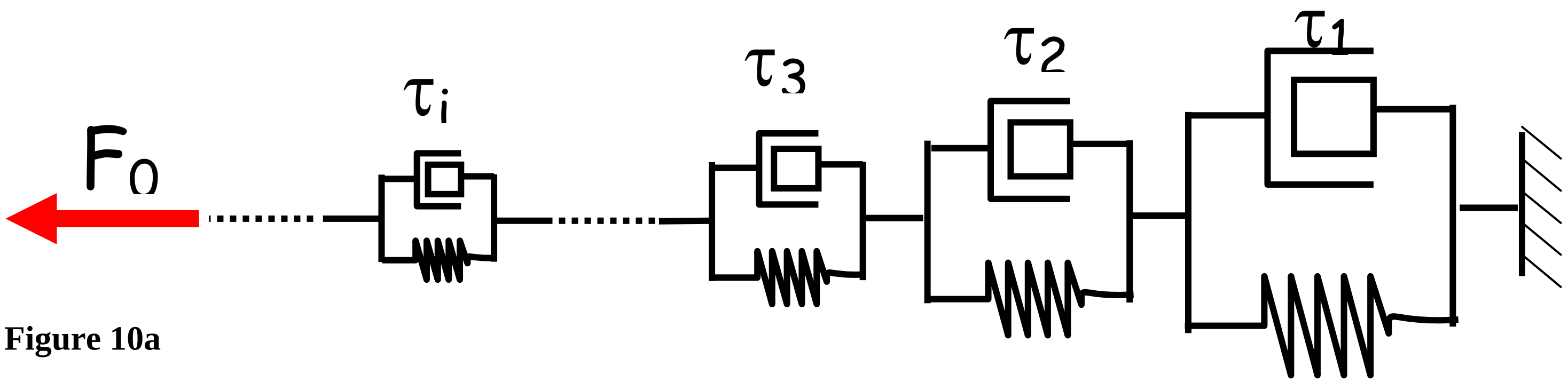

**Figure 10a**

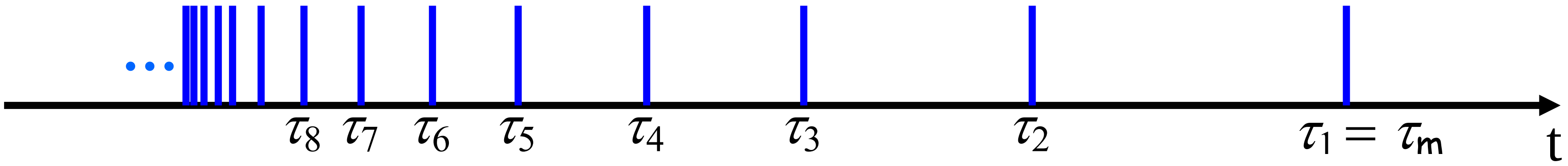

**Figure 10b**

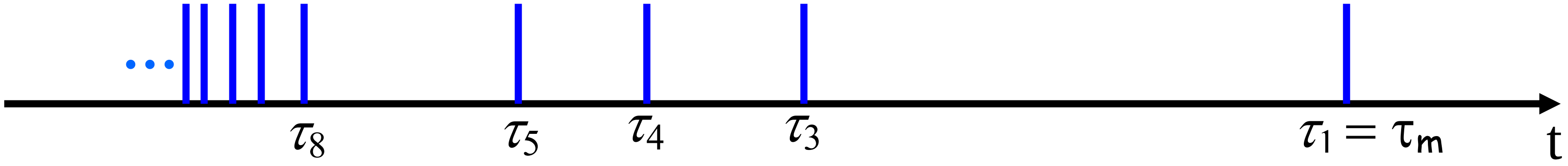

**Figure 10c**

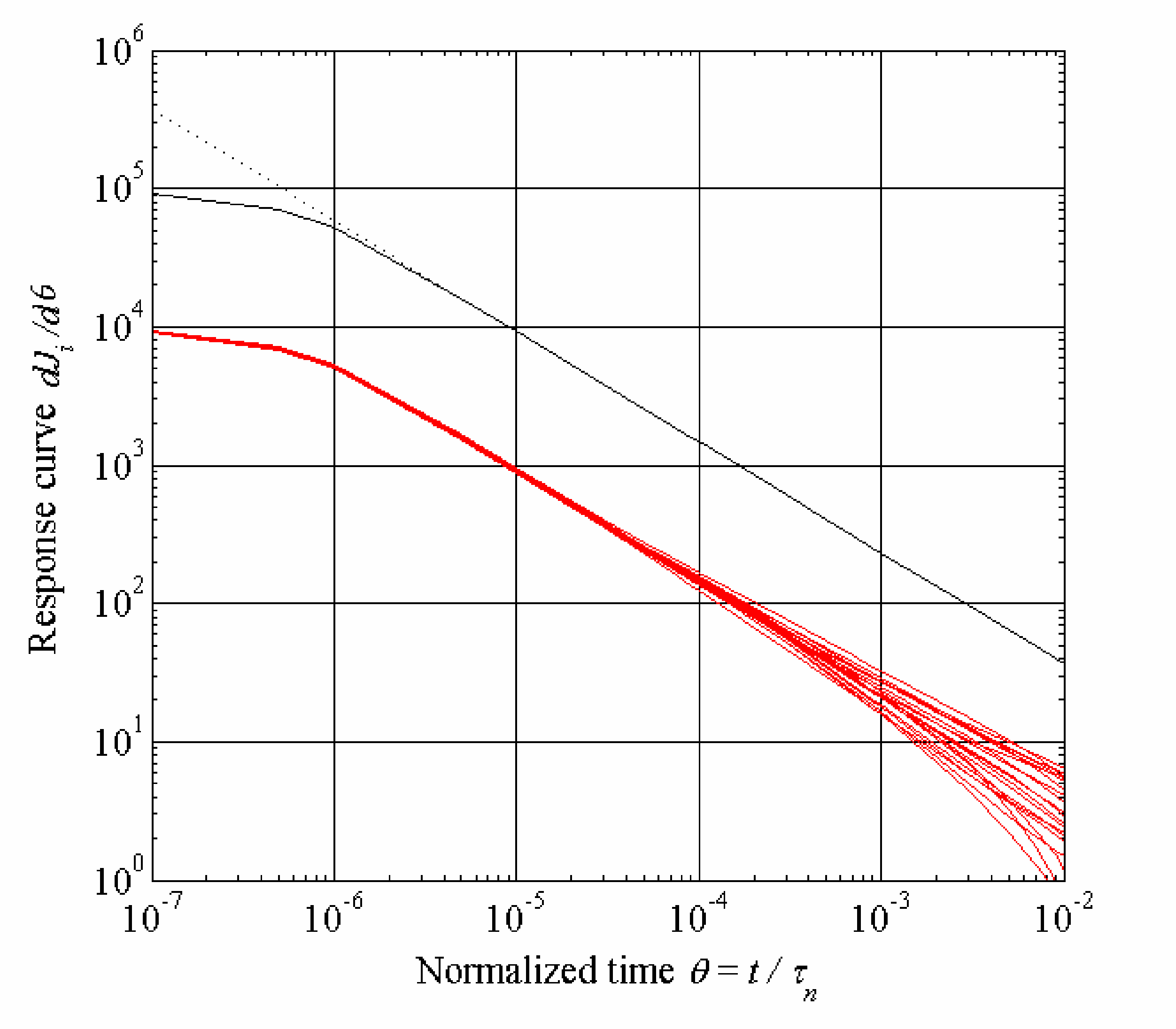

**Figure 11**

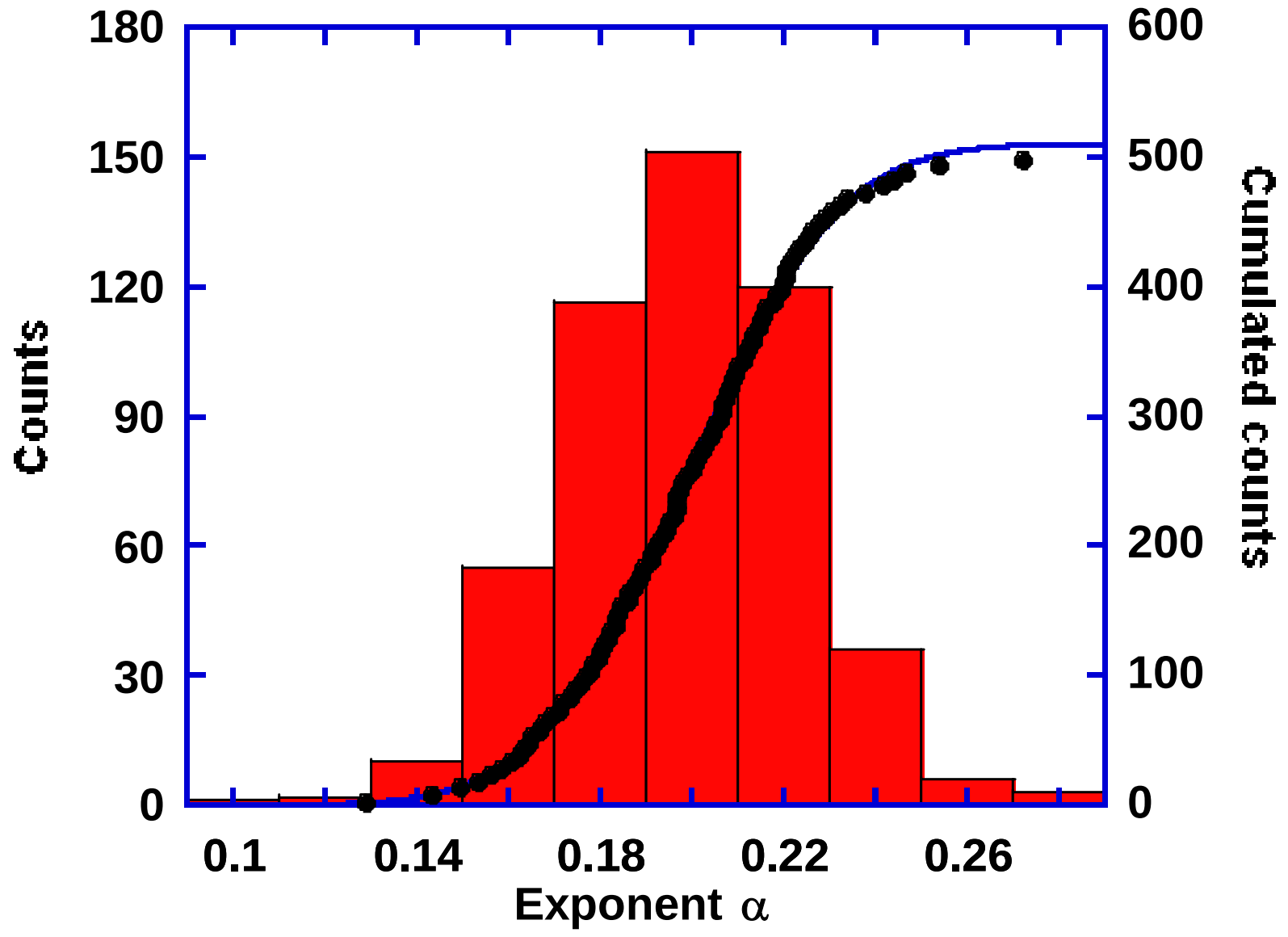

Figure 12a

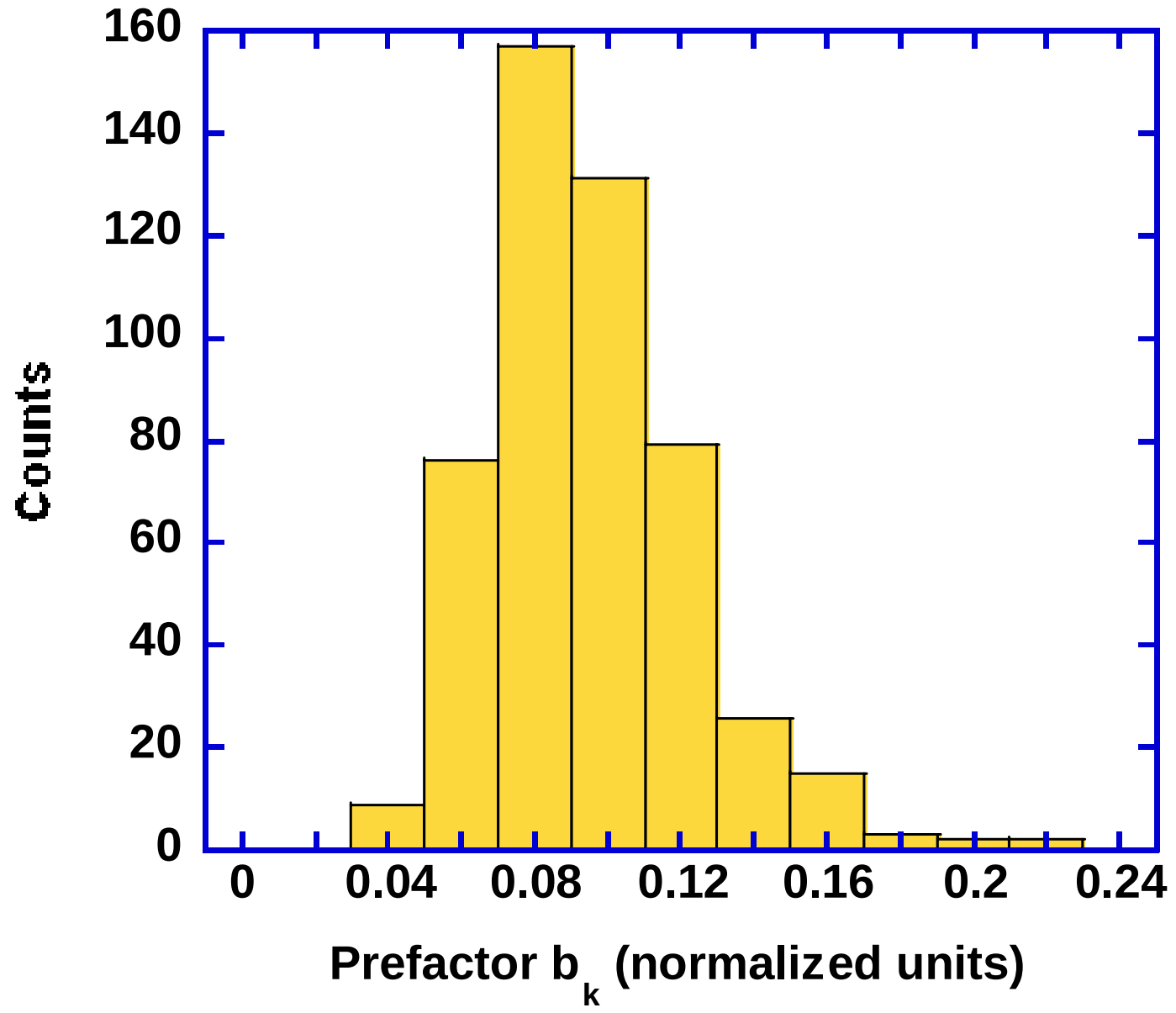

Figure 12b

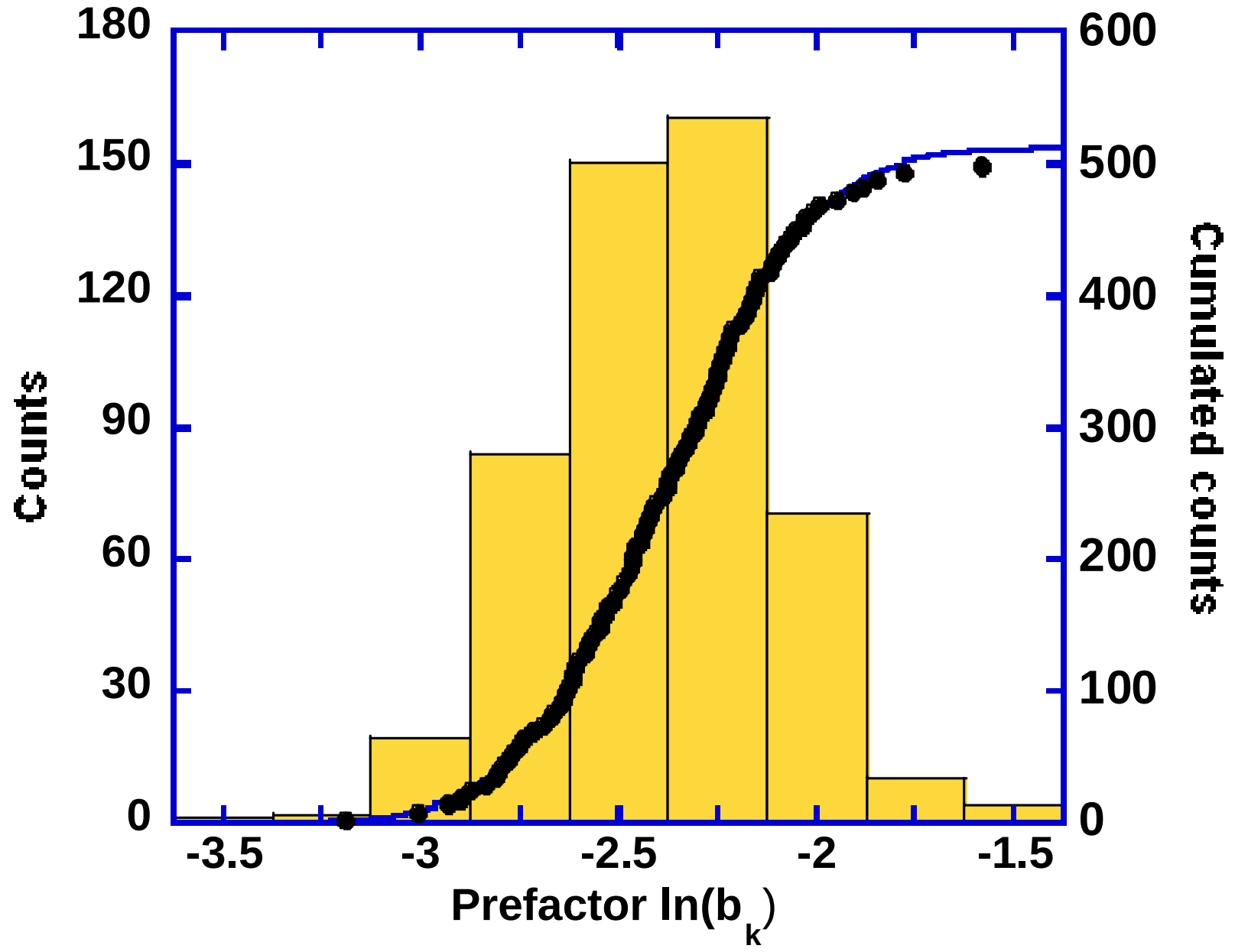

Figure 12c

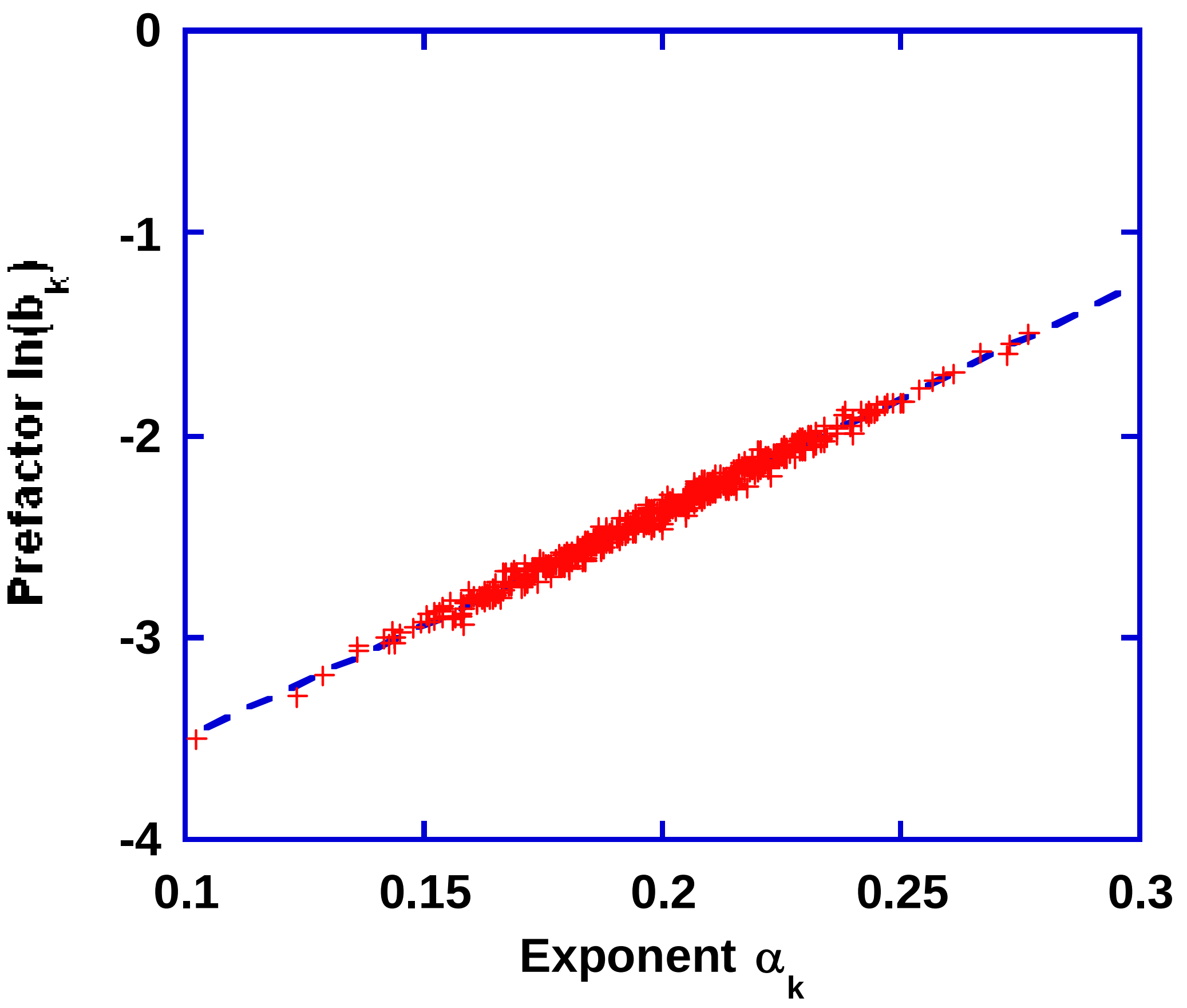

Figure 13

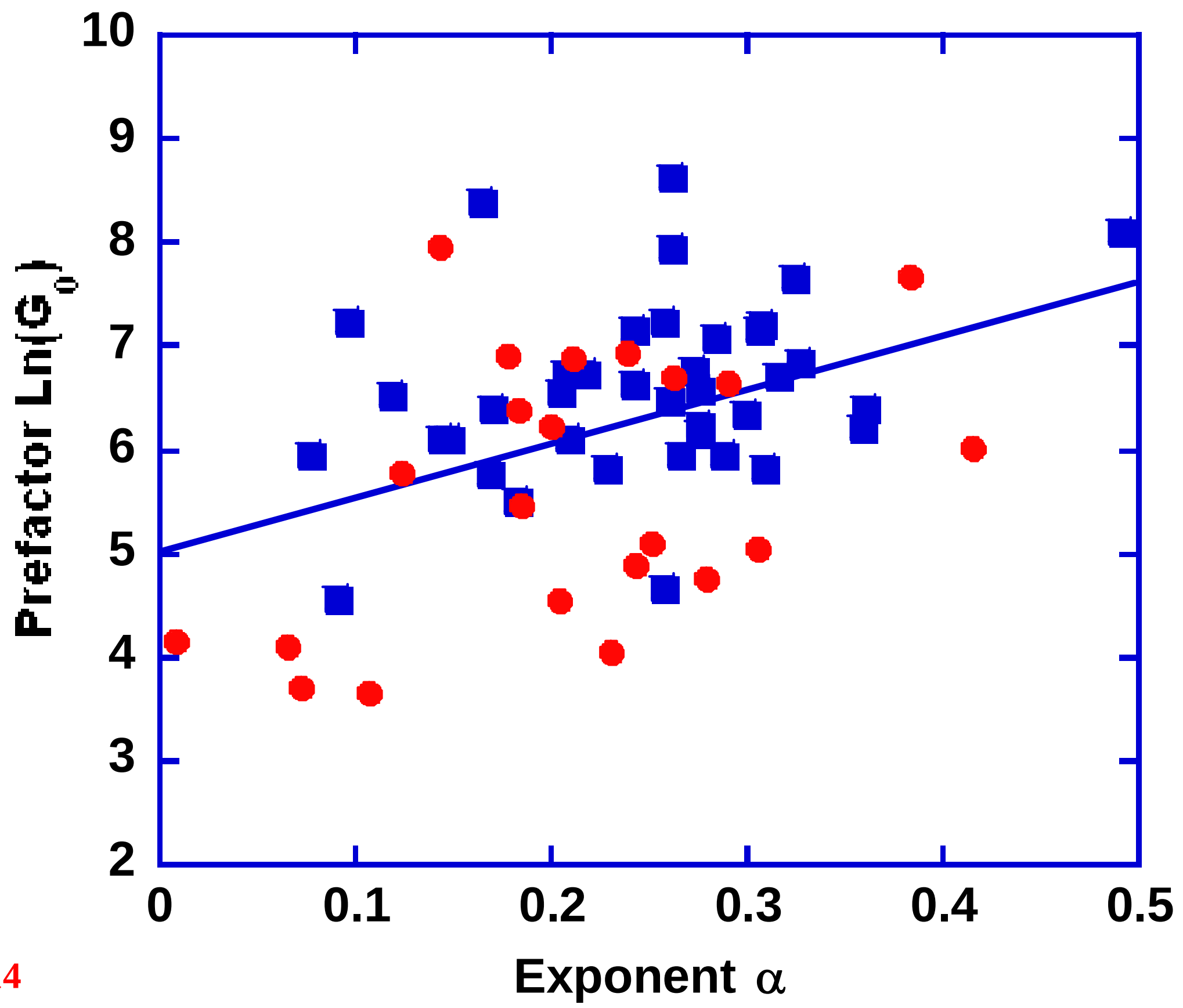

Figure 14